\documentclass[aps,prd,twocolumn,showpacs,preprintnumbers,nofootinbib,
               superscriptaddress]{revtex4}
\usepackage{graphicx, fancybox}
\usepackage{amsmath,amssymb}

\newcommand{\bm}[1]{\mbox{\boldmath$#1$}}
\newcommand{\e}{\mbox{e}}

\begin{document}

\preprint{}

\title{
Quark propagator at finite temperature and finite momentum\\
in quenched lattice QCD
}

\author{Frithjof Karsch}
\email{karsch@bnl.gov}
\affiliation{
Brookhaven National Laboratory, Bldg. 510A, Upton, NY 11973, USA
}
\affiliation{
Fakult\"at f\"ur Physik, Universit\"at Bielefeld, D-33615 Bielefeld, Germany}

\author{Masakiyo Kitazawa}
\email{kitazawa@phys.sci.osaka-u.ac.jp}
\affiliation{
Department of Physics, Osaka University, Toyonaka, Osaka, 560-0043, Japan
}

\begin{abstract}

We present an analysis of the quark spectral function above and
below the critical temperature for deconfinement performed at
zero and non-zero momentum in quenched lattice QCD  
using clover improved Wilson fermions in Landau gauge.
It is found that the temporal quark correlation function
in the deconfined phase near the critical temperature 
is well reproduced by a two-pole ansatz for the spectral 
function. This indicates that excitation modes of 
the quark field have small decay rates.
The bare quark mass and momentum dependence of the 
spectral function is analyzed with this ansatz.
In the chiral limit we find that the quark spectral 
function has two collective modes corresponding to the 
normal and plasmino excitations in the high temperature 
limit. Over a rather wide temperature range in
the deconfined phase the
pole mass of these modes at zero momentum, 
which corresponds to the thermal mass of the quark, 
is approximately proportional to temperature. 
With increasing bare quark masses the plasmino mode 
gradually disappears 
and the spectral function is dominated by 
a single pole. We also discuss 
quasi-particle properties of heavy 
quarks in the deconfined phase.
In the confined phase, it is found that the pole 
ansatz for the spectral function fails completely. 

\end{abstract}

\date{May 24, 2009}

\maketitle


\section{Introduction}
\label{sec:Intro}

The asymptotic freedom of QCD tells us that matter 
at extremely high temperature ($T$) becomes a simple 
thermodynamic system
composed of weakly interacting quarks and gluons. 
Thermodynamic quantities approach the Stefan-Boltzmann 
limit for massless quarks and gluons in the high temperature 
limit.
Deviation from this limit that arise from the 
temperature-dependent running gauge coupling $g$ can 
be calculated perturbatively \cite{BIR,LeBellac}.
The excitation properties of quarks and gluons in this 
region can also be analyzed using perturbative techniques.
The fact that $T$ dominates over all other scales allows 
to adopt the hard-thermal loop (HTL) approximation \cite{HTL},
which enables us to calculate propagators of these fields 
in a gauge independent way.
It is known that in leading order of HTL-resummed
perturbation theory mass gaps arise in the excitation
spectra of these degrees of freedom 
which are proportional to $gT$.
They are called {\it thermal masses} \cite{LeBellac}.
These excitations acquire decay widths owing to 
medium effects, which are proportional to $g^2T$.
At sufficiently high $T$ these widths 
thus are parametrically negligible compared to the thermal 
masses. In addition, a novel excitation, the plasmino, appears
in the quark propagator, which has a minimum in its dispersion
relation at nonzero momentum \cite{plasmino}.
Various discussions addressed the origin of such a peculiar 
mode \cite{Weldon:1989ys,BBS92,Blaizot:1993bb,Peshier:1999dt,
KKN06,Kitazawa:2007ep} and its phenomenological 
consequences \cite{Braaten:1990wp}.

As $T$ is lowered, the gauge coupling $g$ grows 
and perturbation theory eventually breaks down.
At the same time, other scales, which are not 
negligible compared to $T$ emerge, and make the
problem more complicated. This is, in particular,
the case for the 
quark-gluon plasma (QGP) phase near the 
critical temperature of deconfinement, $T_c$, which 
is analyzed experimentally in heavy ion collision
performed currently with the
Relativistic Heavy Ion Collider (RHIC) 
\cite{RHIC}. To analyze the highly non-perturbative
properties of strongly interacting matter in the
temperature range accessible to the RHIC experiments,
a non-perturbative approach to QCD, such as 
lattice QCD Monte Carlo simulation, is needed.

Since the leading order HTL-resummed perturbative calculations 
predict that the decay widths of quarks and gluons 
grow faster than the thermal masses as $T$ is lowered, 
one may na\"ively expect that for temperatures near 
$T_c$ the peaks in spectral functions of these degrees
of freedom disappear.
In other words, a quasi-particle description of these 
excitations would be inappropriate near $T_c$.
There are, however, several arguments supporting 
the existence of quasi-particles with the quantum numbers
of quarks in this region.
For example, quasi-particles have been used successfully 
to describe lattice QCD results on the equation of states 
and susceptibilities \cite{EoS}.
The existence of quark quasi-particles near $T_c$ is 
also suggested by results from lattice simulations on 
baryon number, electric charge and strangeness
fluctuations \cite{fluctuations,newfluct}.
It is also notable that quark number scaling of the 
elliptic flow observed in the RHIC experiments 
\cite{Fries:2003kq} suggests the existence of quark degrees
of freedom in the hot and dense matter created at RHIC.
Direct studies of excitation properties of quarks based on 
first principle lattice calculations may therefore help to 
clarify the physics behind these findings.

After some pioneering work on thermal properties 
of the quark propagator in 
lattice QCD simulations \cite{quark_lattice}, 
in \cite{Karsch:2007wc} the authors of the present
paper investigated the spectral properties of 
quarks at zero momentum for two values of temperature 
in the deconfined phase, $T=1.5T_c$ and $3T_c$, in 
quenched lattice QCD in Landau gauge.
In this work, we use a two-pole ansatz for 
the quark spectral function, $\rho(\omega,\bm{p})$, 
motivated by the structure of $\rho(\omega,\bm{p})$ 
in the high temperature limit. This ansatz includes 
contributions of the normal and plasmino modes and allows
to analyze the importance of thermal widths
of these modes.
It is found that the Euclidean correlation function for quarks 
on the lattice is well reproduced by this ansatz.
The result indicates that excitation modes of quarks 
form sharp peaks in $\rho(\omega,\bm{p})$ even near 
$T_c$. In this temperature range close to $T_c$ it
also is found that the form of $\rho(\omega,\bm{p})$ is
similar to that in the perturbative region. It clearly 
is quite different from that at 
$T=0$ which is given by field strength renormalization 
$Z(p)$ and mass function $M(p)$ \cite{T=0}.

At temperatures close to $T_c$, the appearance of 
scales which are not negligible compared to $T$ 
invalidates the simple picture provided by the HTL 
approximation, in addition to the failure of
the perturbative treatment.  From the analysis of 
simpler models, composed of fermions 
and bosons, it is known that the structure of $\rho(\omega,\bm{p})$ 
has a nontrivial dependence on the masses of fermions 
and bosons when these masses are comparable to $T$ 
\cite{BBS92,KKN06,Kitazawa:2007ep,Harada:2008vk}.
Examples are the spectrum of fermions with a mass $m$ 
in QED or a Yukawa model with a massless boson \cite{BBS92}.
In these models $\rho(\omega,\bm{p})$ is given by the HTL 
approximation and, in particular, receives contributions 
from the plasmino mode whenever $T/m$ is large enough.
On the other hand, $\rho(\omega,\bm{p})$ approaches
a free quark spectral function without the plasmino
mode in the low temperature $T/m\to0$ limit.
As discussed in \cite{BBS92}, and summarized in 
Appendix~\ref{sec:Yukawa} in the present paper,
the numerical result, which has been obtain
 at the one loop order, shows that 
$\rho(\omega,\bm{p})$ changes continuously as 
a function of $T/m$ between these two limits.
In \cite{Karsch:2007wc}, we analyzed the dependence
of $\rho(\omega,\bm{p})$ on the bare quark mass,
$m_0$, for each $T$ and found that the lattice 
result are in accordance with these findings.

The main purpose of the present study is to extend
the analysis of \cite{Karsch:2007wc} to lower 
temperatures, closer to $T_c$,  and to non-zero momentum.
In addition to the temperature values analyzed before,
we performed the simulations at three lower 
temperatures above and below $T_c$.
We show that the results obtained in \cite{Karsch:2007wc} 
do not change qualitatively even at $T=1.25T_c$.
On the other hand, we find that the two-pole 
approximation completely fails below $T_c$,
which indicates that excitations of quark fields 
with a narrow width do not exist in the confined phase.
We also analyze the momentum dependence of 
$\rho(\omega,\bm{p})$ and show that the dispersion
relations of the normal and plasmino modes behave
reasonably at finite momentum.
Furthermore, we discuss the spectral properties 
of the light and charm quarks.

An important aspect of the analysis of the quark 
spectral function is that these spectral functions
directly reflect the symmetries of the thermal system
and are sensitive to their explicit or spontaneous breaking. 
One can, for example, clearly observe the effect of chiral symmetry 
breaking in the scalar channel of the quark propagator.
We show that the behavior of scalar channel is quite 
different between below and above $T_c$;
while the scalar contribution to the spectral function 
becomes vanishingly small in the chiral limit 
above $T_c$, such a behavior is not observed below $T_c$.
The behavior of the quark correlation function
on exceptional configurations \cite{excp.conf} 
is reported in detail in Appendix.~\ref{sec:exceptional}.

This paper is organized as follows.
In the next section, we review the general properties
of the quark spectral function and discuss its structure for 
some special cases.
The setup of the numerical simulation is summarized 
in Sec.~\ref{sec:setup}.
We then discuss the numerical results for the spectral 
function for $T>T_c$ in Sec.~\ref{sec:T>Tc} and Sec.\ref{sec:p}.
In Sec.~\ref{sec:T>Tc}, we consider the bare quark mass 
dependence of the spectral function at zero momentum.
This analysis is extended to finite momenta in Sec.~\ref{sec:p}.
Section~\ref{sec:T<Tc} is devoted to a 
discussion of the quark correlation function below $T_c$.
We give a brief summary in Sec.~\ref{sec:summary}.
In Appendix~\ref{sec:Yukawa}, we review the fermion 
spectral function in a Yukawa model.
In Appendix~\ref{sec:exceptional}, 
the behavior of the quark propagator on exceptional 
configuration is presented.

\section{General properties of quark propagator and spectral function}
\label{sec:propagator}

In this section, we summarize the definition and 
properties of the quark propagator and the quark
spectral function.
While the content of this section may be familiar 
to many readers, this section also serves to introduce 
the notation used in subsequent sections.
We also review the forms of the spectral function 
in some limiting cases which are of relevance in later 
sections.

\subsection{Definitions}

The excitation properties of quarks in a thermal medium 
can be extracted from the imaginary-time (Matsubara) 
quark propagator which is defined as,
\begin{align}
S_{\mu\nu}^{bc}( \tau,\bm{x} ; 0,\bm{y} )
= \langle {\rm T}_\tau \psi_\mu^b( \tau,\bm{x} ) 
\bar\psi_\nu^c ( 0,\bm{y} ) \rangle,
\label{eq:S^ab}
\end{align}
where $\tau$ is the imaginary time restricted to the interval 
$0\le\tau<1/T$. Here ${\rm T}_\tau$ denotes the time-ordering
along the imaginary time, and $\psi_\mu^b(\tau,\bm{x})$ 
is the quark operator, with Greek subscripts denoting the 
Dirac indices, and $b$ and $c$ representing colors.
The thermal average $\langle {\cal O} \rangle$ is defined by
$\langle {\cal O} \rangle = (1/Z) {\rm Tr} [ \e^{-\beta H} {\cal O} ]$,
where the trace is taken over a complete set of quantum states and
$Z={\rm Tr} [ \e^{-\beta H} ]$.
In the following analysis, we use the propagator 
in momentum space,
\begin{align}
S_{\mu\nu}( \tau,\bm{p} )
= \frac1{VN_c} \sum_b \int d^3x d^3y 
\e^{ i {\bf p} \cdot ( {\bf x}-{\bf y} ) }
S_{\mu\nu}^{bb}( \tau,\bm{x} ; 0,\bm{y} ),
\label{eq:S:def}
\end{align}
where $V$ denotes the volume of the system and
$N_c=3$ is the number of colors.
Equation~(\ref{eq:S:def}) is referred to as the 
correlation function in the following.
Since $S_{\mu\nu}(\tau,\bm{p})$ is a gauge covariant 
quantity, gauge-fixing conditions are required 
to obtain non-zero expectation values.
Equation~(\ref{eq:S:def}) satisfies  
anti-periodic boundary condition along the imaginary time,
\begin{align}
S_{\mu\nu}( \tau,\bm{p} )
= - S_{\mu\nu}( \tau-1/T,\bm{p} ).
\label{eq:KMS}
\end{align}

For zero quark chemical potential, the thermal ensemble has
a charge conjugation symmetry.
The thermal average thus satisfies 
$\langle {\cal O} \rangle = \langle {\cal COC} \rangle$,
with ${\cal C}$ representing the unitary operator 
for charge conjugation.
This symmetry leads to the following identity of the 
correlation function
\begin{align}
S_{\mu\nu}( \tau,\bm{p} )
&= \left[ C S^T( -\tau,-\bm{p} ) C^{-1} \right]_{\mu\nu}
\nonumber \\
&= -\left[ C S^T( 1/T-\tau,-\bm{p} ) C^{-1} \right]_{\mu\nu},
\label{eq:S:charge}
\end{align}
where $C$
is the charge conjugation matrix; 
${\cal C}\psi{\cal C} = C \psi^T$.
In the last equality of Eq.~(\ref{eq:S:charge}), 
we used Eq.~(\ref{eq:KMS}).

The Fourier transform of Eq.~(\ref{eq:S:def}) 
with respect to $\tau$ is given by
\begin{align}
S_{\mu\nu}( \tau,\bm{p} )
= T\sum_n \e^{ i\omega_n\tau }
S_{\mu\nu}( i\omega_n, \bm{p} ),
\label{eq:S}
\end{align}
with the Matsubara frequencies for fermions $\omega_n=(2n+1)\pi T$.
The retarded propagator in real time is obtained by
the analytic continuation of Eq.~(\ref{eq:S}) as
$S_{\mu\nu}^{\rm R}( \omega,\bm{p} ) 
= S_{\mu\nu}( i\omega_n, \bm{p} )|_{i\omega_n\to \omega+i\eta}$.

The spectral function is related to the retarded and Matsubara 
propagators as
\begin{align}
\rho_{\mu\nu}( \omega,\bm{p} ) 
&= -(1/\pi){\rm Im} S_{\mu\nu}^{\rm R}( \omega,\bm{p} ) 
\nonumber\\
&= -(1/\pi)\left( S^{\rm R}( \omega,\bm{p} ) 
- \gamma^0 S^{\rm R}( \omega,\bm{p} )^\dagger \gamma^0 \right)_{\mu\nu},
\end{align}
and
\begin{align}
S_{\mu\nu}( i\omega_n, \bm{p} )
= \int d\omega' \frac{ \rho_{\mu\nu}( \omega',\bm{p} ) }
{ i\omega_n-\omega' }.
\label{eq:S-rho} 
\end{align}
From Eqs.~(\ref{eq:S-rho}) and (\ref{eq:S}),
one obtains the relation between $\rho_{\mu\nu}(\omega,\bm{p})$ 
and $S_{\mu\nu}(\tau,\bm{p})$;
\begin{align}
S_{\mu\nu}( \tau,\bm{p} )
= \int_{-\infty}^\infty d\omega
\frac{ \e^{ (1/2 -\tau T) \omega/T }}{ \e^{\omega/2T} + \e^{-\omega/2T} }
\rho_{\mu\nu}( \omega,\bm{p} ).
\label{eq:Stau-rho}
\end{align}
Using the charge conjugation symmetry,
one can show that $\rho_{\mu\nu}(\omega,\bm{p})$ obeys
the following relation,
\begin{align}
\rho_{\mu\nu} (\omega,\bm{p})
= [ C \rho^T(-\omega,-\bm{p}) C ]_{\mu\nu}.
\label{eq:rho:charge}
\end{align}

\subsection{Dirac structure}
\label{sec:Dirac}

Owing to parity and rotational symmetries,
the Dirac structure of the quark spectral function 
at finite temperature is in general decomposed as
\begin{align}
\lefteqn{ \rho_{\mu\nu}( \omega, \bm{p} )}
\nonumber \\
&= \rho_0( \omega,p ) (\gamma^0)_{\mu\nu}
- \rho_{\rm v}( \omega,p ) (\hat{\bm{p}}\cdot\bm{\gamma})_{\mu\nu}
+ \rho_{\rm s}( \omega,p ) \bm{1}_{\mu\nu},
\label{eq:rho_0vs}
\end{align}
where $p=|\bm{p}|$ and $\hat{\bm{p}}=\bm{p}/p$.
Here, 
\begin{align}
\rho_0( \omega,p ) 
&= {\rm Tr}_{\rm D} [ \rho( \omega,\bm{p} ) \gamma^0 ]/4,
\label{eq:rho_0} \\
\rho_{\rm v}( \omega,p ) 
&= {\rm Tr}_{\rm D} [ \rho( \omega,\bm{p} ) \hat{\bm{p}}\cdot\bm{\gamma} ]/4,
\label{eq:rho_v} \\
\rho_{\rm s}( \omega,p ) &= {\rm Tr}_{\rm D} [\rho( \omega,\bm{p} )]/4,
\label{eq:rho_s}
\end{align}
with ${\rm Tr}_{\rm D}$ denoting the trace over Dirac indices.
Using Eq.~(\ref{eq:rho:charge}), it is shown that 
$\rho_0( \omega,p )$ is an even function and 
$\rho_{\rm v,s}( \omega,p )$ are odd functions of $\omega$;
\begin{align}
\rho_0( \omega,p ) &= \rho_0( -\omega,p ),
\label{eq:rho_02} \\
\rho_{\rm v}( \omega,p ) &= -\rho_{\rm v}( -\omega,p ),
\label{eq:rho_v2} \\
\rho_{\rm s}( \omega,p ) &= -\rho_{\rm s}( -\omega,p ).
\label{eq:rho_s2}
\end{align}

In some special cases, the Dirac structure of 
the spectral function can also be decomposed 
by using projection operators \cite{Blaizot:1993bb,decomp}.
Two such examples, which are of relevance in the subsequent 
sections, are the spectral function at zero 
momentum and that having chiral symmetry.
For the former case, 
$\rho_{\rm v}( \omega,p )$ vanishes and 
$\rho( \omega,\bm{p}=\bm{0} )$ can be decomposed
with the projection operators 
$L_\pm = ( 1 \pm \gamma^0 )/2$ as 
\begin{align}
\rho( \omega, \bm{0} )
= \rho^{\rm M}_+( \omega ) L_+ \gamma^0 
+ \rho^{\rm M}_-( \omega ) L_- \gamma^0 ,
\label{eq:rho^M}
\end{align}
where
\begin{align}
\rho^{\rm M}_\pm( \omega ) 
&= \frac{1}{2} {\rm Tr}_{\rm D} [ \rho( \omega,\bm{0} ) \gamma^0 L_\pm ]
\nonumber \\
&= \rho_0( \omega,0 ) \pm \rho_{\rm s}( \omega,0 ).
\label{eq:rho^M_pm}
\end{align}
We note that gamma matrices ($\gamma^0$) in Eq.~(\ref{eq:rho^M}) 
come directly from the definition of the quark propagator;
$S(\tau) 
= \langle {\rm T}_\tau \psi(\tau) \bar\psi(0) \rangle
= \langle {\rm T}_\tau \psi(\tau) \psi^\dagger(0) \rangle \gamma^0$.

If the system is chirally symmetric, the quark 
propagator must anti-commute with $\gamma_5$, 
and thus $\rho_{\rm s}( \omega,p )$ vanishes 
in Eq.~(\ref{eq:rho_0vs}).
In this case, $\rho( \omega,\bm{p} )$ can be decomposed
using the projection operators 
$P_\pm (\bm{p})= ( 1 \pm \gamma^0\hat{\bm{p}}\cdot\bm{\gamma} )/2$
as 
\begin{align}
\rho( \omega,\bm{p} )
= \rho^{\rm P}_+( \omega,p ) P_+(\bm{p}) \gamma^0
+ \rho^{\rm P}_-( \omega,p ) P_-(\bm{p}) \gamma^0 ,
\label{eq:rho^P}
\end{align}
where
\begin{align}
\rho^{\rm P}_\pm( \omega,p ) 
&= \frac{1}{2} {\rm Tr}_{\rm D} [ \rho( \omega,\bm{p} ) \gamma^0 P_\pm(\bm{p}) ]
\nonumber \\
&= \rho_0( \omega,p ) \pm \rho_{\rm v}( \omega,p ) .
\label{eq:rho^P_pm}
\end{align}

We note that in general $\rho^{\rm M}_\pm(\omega)$ and 
$\rho^{\rm P}_\pm( \omega,p )$ are
neither even nor odd functions. 
Instead, the charge conjugation symmetry, 
Eq.~(\ref{eq:rho:charge}), requires the following 
relations for $\rho^{\rm M}_\pm(\omega)$ and 
$\rho^{\rm P}_\pm( \omega,p )$;
\begin{align}
\rho^{\rm M}_\pm(\omega) &= \rho^{\rm M}_\mp(-\omega),
\label{eq:rhoMC} \\
\rho^{\rm P}_\pm( \omega,p ) &= \rho^{\rm P}_\mp(-\omega,p ).
\label{eq:rhoPC}
\end{align}

Finally, the chirally symmetric 
spectral function  
at zero momentum is simply written as 
$\rho( \omega,\bm{0} ) = \rho_0(\omega,0)\gamma^0$.
This function can be decomposed into the forms 
given by Eq.~(\ref{eq:rho^M}) as well as Eq.~(\ref{eq:rho^P}) 
with $\rho^{\rm M}_\pm( \omega ) = \rho^{\rm P}_\pm( \omega,0 ) 
= \rho_0(\omega,0) $.
Only in this case do $\rho^{\rm M}_\pm( \omega )$ and 
$\rho^{\rm P}_\pm( \omega,0 )$ become even functions 
of $\omega$.

The spectral functions, $\rho^{\rm M}_\pm(\omega)$ and 
$\rho^{\rm P}_\pm( \omega,p )$, which arise in the
decomposition of the spectral function $\rho(\omega,p)$, 
represent excitations 
having definite quantum numbers corresponding to 
each projection.
Therefore, the excitation properties of quarks 
are more apparent in these channels rather than 
in Eqs.~(\ref{eq:rho_0})-(\ref{eq:rho_s}).
Moreover, using the definition of the spectral 
function, one can prove that they are non-negative,
$\rho^{\rm M}_\pm(\omega)\ge 0$ and 
$\rho^{\rm P}_\pm(\omega,p)\ge 0$.
In the analyses presented in later sections, 
we therefore mainly refer to $\rho_\pm^{\rm M}(\omega)$ 
and $\rho_\pm^{\rm P}(\omega,0)$ instead of 
Eqs.~(\ref{eq:rho_0})-(\ref{eq:rho_s}).
One can also show that $\rho_0(\omega,p)\ge0$,
while the signatures of $\rho_{\rm v}(\omega,p)$ and 
$\rho_{\rm s}(\omega,p)$ are not determined from the definition.

The correlation function $S_{\mu\nu}(\tau,\bm{p})$ is also 
decomposed similarly to Eq.~(\ref{eq:rho_0vs}),
\begin{align}
\lefteqn{S_{\mu\nu} (\tau,\bm{p})}
\nonumber \\
&= S_0(\tau,p) (\gamma^0)_{\mu\nu}
- S_{\rm v}( \tau,p ) (\hat{\bm{p}}\cdot\bm{\gamma})_{\mu\nu}
+ S_{\rm s}( \tau,p ) \bm{1}_{\mu\nu}.
\label{eq:S_0vs}
\end{align}
Using the charge conjugation symmetry, one can show that
$S_0(\tau,p)=S_0(1/T-\tau,p)$, 
$S_{\rm v,s}(\tau,p)=-S_{\rm v,s}(1/T-\tau,p)$.
For zero momentum and in the chiral limit, 
the correlation function reduces to 
\begin{align}
S (\tau,\bm{0}) \gamma^0
&= S^{\rm M}_+( \tau ) L_+ 
+ S^{\rm M}_-( \tau ) L_- ,
\label{eq:S^M}
\\
S (\tau,\bm{p}) \gamma^0
&= S^{\rm P}_+( \tau,p ) P_+(\bm{p}) 
 + S^{\rm P}_-( \tau,p ) P_-(\bm{p}) ,
\label{eq:S^P}
\end{align}
respectively.
The charge conjugation symmetry requires that 
\begin{align}
S^{\rm M}_\pm(\tau)=S^{\rm M}_\mp(1/T-\tau),\quad
S^{\rm P}_\pm(\tau,p)=S^{\rm P}_\mp(1/T-\tau,p),
\label{eq:S_pm}
\end{align}
while $S^{\rm M}_\pm(\tau)$ and $S^{\rm P}_\pm(\tau,p)$
are neither symmetric nor anti-symmetric.
Only if the system is chirally symmetric, 
$S^{\rm M}_+(\tau)$ becomes a symmetric function,
\begin{align}
S^{\rm M}_+(\tau)=S^{\rm M}_+(1/T-\tau).
\label{eq:S_chiral}
\end{align}

\subsection{Spectral functions in some special cases}
\label{sec:rho:special}

\subsubsection{Free quarks}
\label{sec:rho:free}

The retarded propagator of a free quark with Dirac 
mass $m$ is given by
\begin{align}
S^{\rm R}_{\rm free}( \omega,\bm{p} )
= \frac1{ P \hspace{-0.6em} / - m }
=\frac{ \Lambda_+(\bm{p};m) \gamma^0 }{ \omega+i\eta - E_p }
+ \frac{ \Lambda_-(\bm{p};m) \gamma^0 }{ \omega+i\eta + E_p },
\end{align}
where $P_\mu=(\omega+i\eta,\bm{p})$, and 
$\Lambda_\pm (\bm{p};m)
= ( E_p \pm \gamma^0 \bm{p}\cdot\bm{\gamma} )/ (2E_p) $
are the projection operators for the free quark with
$E_p=\sqrt{ \bm{p}^2 + m^2 }$.
The corresponding spectral function is given by
$\rho( \omega,\bm{p} )
= \rho^{\rm free}_+( \omega,p ) \Lambda_+(\bm{p};m) \gamma^0
+ \rho^{\rm free}_-( \omega,p ) \Lambda_-(\bm{p};m) \gamma^0 $ 
with
\begin{align}
\rho^{\rm free}_\pm( \omega,p )
= \delta( \omega \mp E_p ).
\label{eq:rho_free}
\end{align}
The projection operators $\Lambda_\pm(\bm{p};m)$ satisfy
$\Lambda_\pm(\bm{0};m)=L_\pm$ and 
$\Lambda_\pm(\bm{p};0)=P_\pm(\bm{p})$.
Therefore, $\rho^{\rm M}_\pm(\omega)$ 
of the free quark with zero momentum reads 
\begin{align}
\rho^{\rm M}_\pm(\omega)   &= \delta(\omega\mp m), 
\label{eq:rho^M_free}
\end{align}
and $\rho^{\rm P}_\pm(\omega,p)$ with zero quark mass
\begin{align}
\rho^{\rm P}_\pm(\omega,p) &= \delta(\omega\mp p).
\label{eq:rho^P_free}
\end{align}

\subsubsection{High temperature limit}
\label{sec:rho:HTL}

The quark propagator at asymptotically high temperature 
can be calculated perturbatively.
The validity of HTL approximation allows to obtain 
gauge independent result within this approach \cite{HTL}.
The quark propagator at leading order in perturbation 
is given by 
\begin{align}
S^{\rm R}_{\rm HTL}( \omega,\bm{p} )
&= \left[ P \hspace{-0.6em} / 
- \Sigma_{\rm HTL} ( \omega,\bm{p} ) \right]^{-1},
\label{eq:S_HTL}
\end{align}
where
\begin{align}
\lefteqn{ \Sigma_{\rm HTL} ( \omega,\bm{p} ) }
\nonumber \\
&= \frac {m_T^2}p Q_0 \left( \frac\omega{p} \right) \gamma^0 
+ \frac {m_T^2}p  \left( 1 - \frac{\omega}{p} Q_0 \left( \frac\omega{p} \right)
\right) \hat{\bm{p}} \cdot \bm{\gamma},
\label{eq:SigmaHTL}
\end{align}
is the quark self-energy with thermal mass 
$m_T ^2  = (1/6) g^2 T^2$ and
$ Q_0 = (1/2) \ln (x+1)/(x-1)$ \cite{LeBellac}.
Since Eq.~(\ref{eq:S_HTL}) is chirally symmetric, 
$S^{\rm R}_{\rm HTL}(\omega,\bm{p})$ and the corresponding 
spectral function can be decomposed using the 
projections operators $P_\pm(\bm{p})$.
The spectral functions $\rho^{\rm P}_\pm(\omega,\bm{p})$ 
then read
\begin{align}
\rho_\pm^{\rm P} ( \omega,p )
=& Z_{\rm N}(p) \delta( \omega \mp E_{\rm N}(p) )
+ Z_{\rm P}(p) \delta( \omega \pm E_{\rm P}(p) )
\nonumber \\
&+ \rho_{\rm cont.}( \pm\omega,p ),
\label{eq:rho^P_HTL}
\end{align}
where $\rho_{\rm cont.}( \omega,p )$ represents the 
contribution of the continuum
taking non-zero values in the space-like region.
$\rho_+^{\rm P} ( \omega,p )$ has two poles in the time-like 
region at $E_{\rm N}(p)>0$ and $E_{\rm P}(p)>0$,
which are called normal and (anti-)plasmino modes,
respectively \cite{LeBellac}.
For zero momentum, $E_{\rm N}(0)=E_{\rm P}(0)=m_T$ and 
the residues satisfy $Z_{\rm N}(p) = Z_{\rm P}(p) = 0.5$.
The spectral functions for zero momentum thus are given by,
\begin{align}
\rho_\pm^{\rm P} ( \omega,0 )
&= \rho_\pm^{\rm M} ( \omega ) = \rho_0(\omega,0)
\nonumber \\
&= \frac12 \left[ \delta( \omega - m_T )
+ \delta( \omega + m_T ) \right]. 
\label{eq:rho^M_HTL}
\end{align}

\subsubsection{Effect of a non-zero Dirac mass}
\label{sec:rho:Yukawa}

In the derivation of Eq.~(\ref{eq:S_HTL}), it is 
assumed that not only $g\ll 1$ but also $T$ dominates 
over all other scales, where the latter condition is 
required for the validity of the HTL approximation.
If $T$ is not large enough compared to other scales, 
the effect of the latter shows up which leads to 
modifications of the 
form of the quark propagator even if perturbation
theory is still valid.
An example for such a scale is the Dirac mass of the 
quark, $m$.
The effect of $m$ has been first investigated in 
\cite{BBS92} for the case of QED and for a Yukawa model 
composed of a massive fermion and a massless boson.
In these models, the fermion spectral functions for 
zero momentum, $\rho^{\rm M}_\pm(\omega)$, take simple 
forms in the massless and infinite mass limits:
For $m/T\to0$, $T$ dominates over other scales 
and $\rho^{\rm M}_\pm(\omega)$ should approach 
Eq.~(\ref{eq:rho^M_HTL}),
\begin{align}
\rho_\pm^{\rm M} ( \omega )
\simeq \frac12 \left[ \delta( \omega - m_T )
+ \delta( \omega + m_T ) \right].
\label{eq:rho_Yukawa1}
\end{align}
However, in the opposite limit, $m/T\to\infty$, the 
Dirac mass dominates over $T$ and the spectral 
function approaches that of a free quark,
\begin{align}
\rho_\pm^{\rm M} ( \omega )
\simeq \delta( \omega \mp m ).
\label{eq:rho_Yukawa2}
\end{align}

By comparing Eqs.~(\ref{eq:rho_Yukawa1}) and 
(\ref{eq:rho_Yukawa2}), it is obvious that the number 
of poles in $\rho^{\rm M}_\pm(\omega)$ is different 
in these limits.
The analysis performed in \cite{BBS92} in the one-loop 
approximation showed that these two limits are 
nevertheless connected continuously;
as $m/T$ becomes smaller, the peak corresponding 
to the plasmino gradually manifests itself in 
$\rho^{\rm M}_\pm(\omega)$.
In Appendix~\ref{sec:Yukawa}, the numerical results 
for this feature in the Yukawa model and details of 
the formalism are summarized.
Also in QCD, if the temperature is high enough so that 
the one-loop approximation for the quark self-energy 
is valid, we find the same limiting behavior as in
Eqs.~(\ref{eq:rho_Yukawa1}) and (\ref{eq:rho_Yukawa2}).
At intermediate values of $m/T$ a
similar behavior of $\rho^{\rm M}_\pm(\omega)$  as
found in the model calculations is therfore
also expected.
Using lattice simulations, we will show in the following
that the two limiting forms of the spectral function are 
observed even in the 
non-perturbative region near $T_c$ \cite{Karsch:2007wc} 
in Sec.~\ref{sec:T>Tc}.

\section{Simulation setup}
\label{sec:setup}

In this study, we analyze the quark correlation function, 
Eq.~(\ref{eq:S:def}), using lattice QCD simulations in the 
quenched approximation, where vacuum excitations of the 
quark--anti-quark pairs are neglected.
For the lattice fermion, we use non-perturbatively 
improved clover Wilson fermions 
\cite{Sheikholeslami:1985ij,Luscher:1996jn}.

We use gauge field ensembles which have been generated and 
used previously by the Bielefeld group to study screening 
masses and spectral functions \cite{Bielefeld,Datta:2003ww}.
The simulation parameters are summarized in 
Table~\ref{table:param}.
We calculate the fermion correlation function for five 
values of the temperature, three of which are above $T_c$ 
and the others are below $T_c$.
The simulation for $T>T_c$ is performed on lattices 
of three different volumina, $N_\sigma^3 \times N_\tau$, 
and lattice spacing, $a$, in order to check the dependence 
of the numerical result on volume and lattice spacing.
A column labeled $c_{\rm SW}$ in Table~\ref{table:param} 
gives the parameter for the clover coefficient.
For configurations above $T_c$, we have checked that the 
average of the Polyakov loop on every configuration 
is closest to the $Z(3)$ root on the real axis.

To estimate the value of the lattice spacing at which
our calculations for a given $T/T_c$ have been performed 
we use $T_c\simeq$~300MeV. This results from determinations
of $T_c$ in units of the square root of the string tension,
$T_c/\sqrt{\sigma}\simeq 0.64$ \cite{Teper,sigma,Namekawa:2001ih}
and a value for the string tension, $\sqrt{\sigma}\simeq 465$~MeV,
which is extracted from studies of the heavy quark
potential in QCD with light quarks \cite{asqtad_pot,our_eos}.
We note that the resulting estimate for the lattice cut-off
has to rely on a physical scale that needs to be taken from
a physical, {\it i.e.} unquenched QCD, calculation.

\begin{table}
\begin{center}
\begin{tabular}{ccccccccc}
\hline
\hline

$T/T_c$ & $N_\tau$ & $N_\sigma$ & $\beta$ & $c_{\rm SW}$ 
& $a$[fm] & $N_{\rm conf}$ & $N_{\rm excp}$ \\
\hline
$3$     & $16$ & $64$ & $7.457$ & $1.3389$ & $0.014$ & $51$ & $0$\\
        & $16$ & $48$ & $7.457$ & $1.3389$ & $0.014$ & $51$ & $0$\\
        & $12$ & $48$ & $7.192$ & $1.3550$ & $0.019$ & $51$ & $0$\\
\hline
$1.5$   & $16$ & $64$ & $6.872$ & $1.4125$ & $0.027$ & $44$ & $7$\\
        & $16$ & $48$ & $6.872$ & $1.4125$ & $0.027$ & $51$ & $0$\\
        & $12$ & $48$ & $6.640$ & $1.4579$ & $0.037$ & $51$ & $3$\\
\hline
$1.25$  & $16$ & $64$ & $6.721$ & $1.4404$ & $0.033$ & $48$ & $31$\\
        & $16$ & $48$ & $6.721$ & $1.4404$ & $0.033$ & $58$ & $0$\\
\hline
$0.93$  & $16$ & $48$ & $6.499$ & $1.4579$ & $0.038$ & $50$ & $0$\\
\hline
$0.55$  & $16$ & $48$ & $6.136$ & $1.6530$ & $0.075$ & $60$ & $1$\\
\hline
\hline
\end{tabular}
\end{center}
\caption{
Simulation parameters on lattices of size $N_\sigma^3\times N_\tau$.
The last column labeled $N_{\rm excp}$ gives the number of 
exceptional configurations (see text).
}
\label{table:param}
\end{table}

Quark propagators have been calculated after fixing each 
gauge field configuration to Landau gauge, $\partial_\mu A^\mu=0$. 
For this we used conventional and stochastic minimization 
algorithms with a stopping criterion, 
$(1/3){\rm tr}|\partial_\mu A^\mu|^2 <\alpha$
with $\alpha=10^{-11}$.
By comparing the correlation functions calculated with 
stopping criteria $\alpha=10^{-11}$, $10^{-9}$ and $10^{-7}$,
we have checked that the numerical result converges well
at $\alpha=10^{-11}$.

In the Wilson fermion formulation the bare quark mass, $m$, 
is related to the  hopping parameter $\kappa$.
A na\"ive formula for this relation is 
\begin{align}
m_0 = \frac1{2a} \left( \frac1\kappa - \frac1{\kappa_c} \right) \; ,
\label{eq:m_0}
\end{align}
where $\kappa_c$ denotes the critical hopping parameter 
corresponding to the chiral limit.
For temperatures above $T_c$, we determine $\kappa_c$ 
from the $\kappa$ dependence of the quark propagator,
as will be described more precisely in Sec.~\ref{sec:T>Tc}.
The pole of the free Wilson fermion propagator at zero 
momentum, on the other hand, is given by
\begin{align}
m_p = \frac1a \log( 1 + am_0 ).
\label{eq:m_p}
\end{align}
In the following, we use Eq.~(\ref{eq:m_p}) to define the 
bare quark mass, since we found that the $a$ dependence of 
the quark spectral function at large bare quark mass is 
smaller with the definition Eq.~(\ref{eq:m_p}) rather than 
Eq.~(\ref{eq:m_0}).
The choice of the definitions of the bare quark 
mass, however, does not alter almost all discussions 
in this paper anyway, since our discussions never 
use the precise values of the bare quark mass.

For temperatures $T/T_c\le1.5$ and values of
the hopping parameters close to $\kappa_c$
we observe on some gauge field configurations 
an anomalous behavior of the quark propagator.
The appearance of such exceptional configurations 
in calculations with light quarks in quenched QCD
is a well-known problem in calculations with 
Wilson fermions \cite{excp.conf}.
We discuss the behavior of the quark correlation 
function on these configurations in 
Appendix~\ref{sec:exceptional}.
As summarized there, the behavior of the quark propagator 
on these configurations is clearly unphysical,
and it is easy to introduce a reasonable criterion 
to distinguish them from the normal ones.
The number of configurations identified to be exceptional 
is given in the last column of Table~\ref{table:param} 
labeled $N_{\rm excp}$.
We excluded these configurations from our analysis.
The number of configurations analyzed, $N_{\rm conf}$,
which does not include the exceptional ones, is 
also given in Table~\ref{table:param}.
One sees from this table that the number of the 
exceptional configurations tends to increase as 
$T$ is lowered and as spatial volume is larger.
In particular, we did not observe any exceptional 
configurations for $T/T_c=3$.
On lattices for $64^3\times16$ and $T=1.25T_c$, 
on the other hand, almost $40\%$ configurations 
are identified to be the exceptional ones.
As discussed in Appendix~\ref{sec:exceptional},
this large number is attributed to a strong 
correlation in the appearance of exceptional 
configurations against the gauge update.
We thus consider that the analysis with remaining 
$60\%$ configurations still makes sense; 
see Appendix~\ref{sec:exceptional}.

The quark correlation function, Eq.~(\ref{eq:S^ab}), is the
inverse of the fermion matrix $K=D \hspace{-0.6em} / - m $.
To evaluate it numerically on the lattice, 
we solve the linear equation
\begin{align}
\psi_{\rm source} = K \psi_{\rm result},
\label{eq:psiKpsi}
\end{align}
for a given source vector $\psi_{\rm source}$.
In this study, we use a wall source with momentum $\bm{p}$
\begin{align}
\psi_{\rm source}(\tau,\bm{x}) 
= \frac1V \delta_{\tau,0} \exp(-i\bm{p}\cdot\bm{x}),
\end{align}
and construct the quark propagator, Eq.~(\ref{eq:S:def}), 
from the solution of Eq.~(\ref{eq:psiKpsi}),
$\psi_{\rm result}=K^{-1}\psi_{\rm source}$, as
\begin{align}
S (\tau,\bm{p}) 
&= \sum_{x} \e^{i{\bf p}\cdot{\bf x}}
\psi_{\rm result}(\tau,\bm{x})
\nonumber \\
&= \frac1V \sum_{x,y} \e^{i{\bf p}\cdot ({\bf x-y})}
S( \tau,\bm{x} ; 0,\bm{y} ),
\label{eq:wall}
\end{align}
where the Dirac and color indices are suppressed 
for simplicity.
The point source $\psi_{\rm source}(\tau,\bm{x}) 
= \delta_{\tau,0}\delta_{{\bf x},{\bf 0}}$, 
on the other hand, is the simplest choice for the 
source term, which leads to a different formula for 
the correlation function,
\begin{align}
S(\tau,\bm{p}) 
= \sum_{x} \e^{i{\bf p}\cdot{\bf x}} \psi_{\rm result} ( \tau,\bm{x} )
= \sum_x \e^{i{\bf p}\cdot{\bf x}}
S( \tau,\bm{x} ; 0,\bm{0} ).
\label{eq:point}
\end{align}
Translational invariance requires that the two definitions 
for $S(\tau,\bm{p})$, Eqs.~(\ref{eq:wall}) and 
(\ref{eq:point}), should give the same result.
We have confirmed that this is indeed the case 
within statistical error.
It is, however, found that the statistical error obtained
with Eq.~(\ref{eq:wall}) is notably smaller than that 
with Eq.~(\ref{eq:point}) while the numerical costs are 
almost the same for both definitions.
The advantage of the wall source becomes more 
prominent on lattices with larger spatial volume.
This behavior is understood intuitively:
In Eq.~(\ref{eq:wall}), the propagators of the quark 
field starting from various points at $\tau=0$ are 
averaged; this does suppress fluctuations arising 
from the local structure of gauge configurations.

In the subsequent sections we limit our analyses to two 
cases; (1) zero momentum correlators with non-zero 
values of the mass, $m_p$, and 
(2) finite momentum correlators in the chiral limit and above $T_c$.
The correlation function for each case is decomposed as 
given in Eqs.~(\ref{eq:S^M}) and (\ref{eq:S^P}), respectively.
In order to reduce the statistical error of 
$S^{\rm M}_+(\tau)$ and $S^{\rm P}_+(\tau)$ 
optimally we make use of their periodicity in Euclidean
time, Eq.~(\ref{eq:S_pm}), and define these correlation 
functions on the lattice, for example $S^{\rm M}_+(\tau)$, as
\begin{align}
S_+^{\rm M}(\tau)_{\rm latt.} 
= \frac12 [ S^{\rm M}_+ (\tau) + S^{\rm M}_- (1/T-\tau) ].
\label{eq:S_latt}
\end{align}

\section{\boldmath Quark propagator above $T_c$ at zero momentum}
\label{sec:T>Tc}

In this section, we analyze the quark spectral function 
above $T_c$ for zero momentum but with finite bare quark mass.
As discussed in Sec.~\ref{sec:propagator}, the quark 
spectral function at zero momentum is decomposed into 
$\rho^{\rm M}_\pm(\omega)$ as in Eq.~(\ref{eq:rho^M_pm}).
In the following, we consider $\rho^{\rm M}_+(\omega)$,
since $\rho^{\rm M}_-(\omega)$ is then immediately 
obtained with Eq.~(\ref{eq:rhoMC}).

\subsection{Lattice correlation function and fitting ansatz}

In order to extract the spectral function 
$\rho^{\rm M}_+(\omega)$ from the lattice correlation 
function using Eq.~(\ref{eq:Stau-rho})
we assume a simple ansatz for the shape of 
$\rho^{\rm M}_+(\omega)$ including few fitting parameters.
For the fitting function, we have tried four ans\"atze,
two of which are single- and two-pole ones,
\begin{align}
\rho^{\rm M}_+(\omega)
&= Z_1 \delta( \omega - E_1 ),
\label{eq:1pole}
\\
\rho^{\rm M}_+(\omega)
&= Z_1 \delta( \omega - E_1 ) + Z_2 \delta( \omega + E_2 ).
\label{eq:2pole}
\end{align}
Here $Z_{1,2}$, and $E_{1,2}>0$, are fitting parameters,
which represent the residues and positions of poles, 
respectively.
The pole at $\omega=-E_2$ in Eq.~(\ref{eq:2pole}) 
corresponds to the plasmino mode at high temperatures,
while the pole at $\omega=E_1$ is the normal one.
We have also used fitting functions that allow for a Gaussian widths,
\begin{align}
\rho^{\rm M}_+(\omega)
&= \frac {Z_1}{\sqrt{\pi}\Gamma_1} \exp \frac{-(\omega-E_1)^2}{\Gamma_1^2},
\label{eq:1gauss}
\\
\rho^{\rm M}_+(\omega)
&= \frac {Z_1}{\sqrt{\pi}\Gamma_1} \exp \frac{-(\omega-E_1)^2}{\Gamma_1^2} 
+  \frac {Z_2}{\sqrt{\pi}\Gamma_2} \exp \frac{-(\omega+E_2)^2}{\Gamma_2^2}, 
\label{eq:2gauss}
\end{align}
where $\Gamma_{1,2}$ are additional fitting parameters
corresponding to the width of each peak.

Comparing the values of $\chi^2/{\rm dof}$ in  
correlated fits based on Eqs. (\ref{eq:1pole}) and 
(\ref{eq:2pole}), we found 
for all parameter sets analyzed in the present study
that $\chi^2/{\rm dof}$ 
in a fit based on Eq.~(\ref{eq:2pole}) is more than two 
orders of magnitude smaller than fits based on the
single pole ansatz, Eq.~(\ref{eq:1pole}). 
The pole corresponding to the plasmino mode at $\omega=-E_2$ 
therefore is an intrinsic feature of the quark propagator
above $T_c$ and is needed to describe the numerical results. 
A single-pole ansatz Eq.~(\ref{eq:1pole}) is clearly ruled out.
Using correlated fits based on Eqs. (\ref{eq:1gauss}) 
and (\ref{eq:2gauss}), the minimal $\chi^2$ always is found
at $\Gamma_1=\Gamma_2=0$, {\it i.e.} the ans\"atze reduce to 
Eqs.~(\ref{eq:1pole}) and (\ref{eq:2pole}), respectively.
The extension to include the Gaussian widths 
therefore does not modify the fit at all.
In the following analysis, we thus use the two-pole
ansatz Eq.~(\ref{eq:2pole}).

Here, we note that the above result on the Gaussian 
widths is obtained in {\it correlated} fit.
We checked that if we use {\it uncorrelated} fits, 
which neglect correlations between different $\tau$'s, 
the Gaussian ans\"atze can improve the $\chi^2/{\rm dof}$ 
especially for large bare quark masses.
The numerical result shows that for
large bare quark masses even the single pole ansatz with
non-zero width, 
Eq.~(\ref{eq:1gauss}), including three parameters can give 
smaller $\chi^2/{\rm dof}$ than the four parameter fit
based on Eq.~(\ref{eq:2pole}). 
This shows that there exist strong correlations between 
different time slices on the lattice, which of course is expected.

\begin{figure}[tbp]
\begin{center}
\includegraphics[width=.49\textwidth]{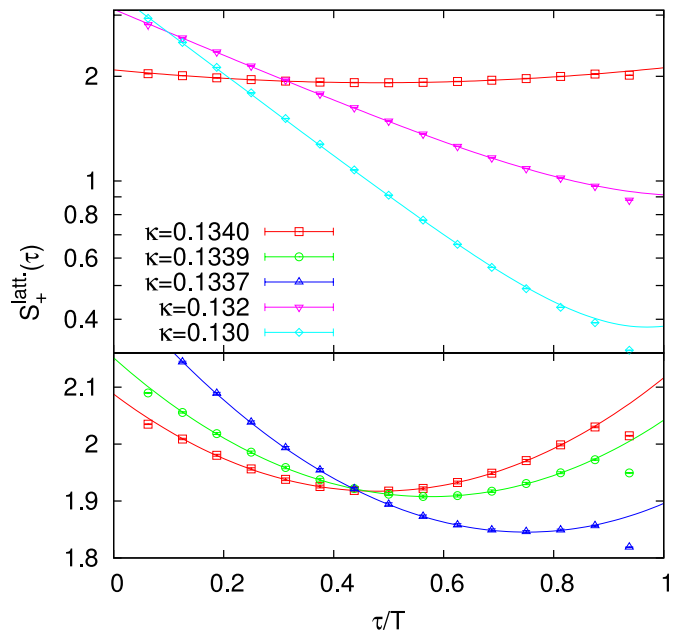}
\caption{
Lattice correlation function $S^{\rm M}_+(\tau)_{\rm latt.}$
at $T=3T_c$ for the lattice of size $64^3\times16$
with various values of $\kappa$.
The solid lines represent the fitting result with the 
two-pole ansatz, Eq.~(\ref{eq:2pole}). Note that the upper panel
shows correlation functions for the heavier quarks on a logarithmic
scale and also includes the correlation function for the lightest
quark mass, which also is shown again in the lower panel on a linear
scale.
}
\label{fig:S}
\end{center}
\end{figure}

\begin{figure}[tbp]
\begin{center}
\includegraphics[width=.49\textwidth]{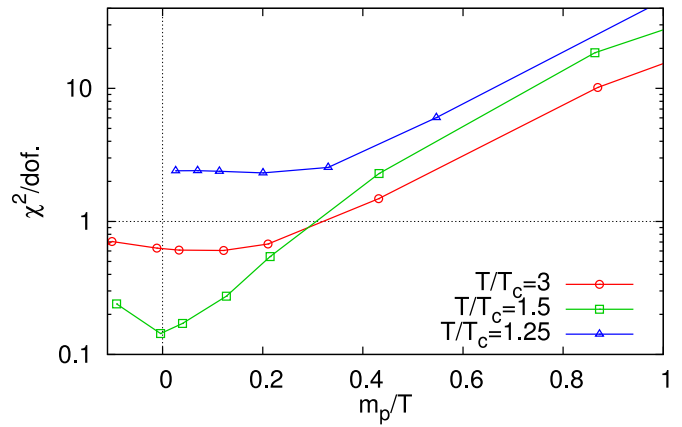}
\caption{
Bare quark mass dependence of the $\chi^2/{\rm dof}$
obtained for two-pole fits 
at $T/T_c=1.25$, $1.5$ and $3$ on lattices of size $64^3\times16$.
}
\label{fig:chi2_p0}
\end{center}
\end{figure}

In Fig.~\ref{fig:S}, we show the numerical results
for $S^{\rm M}_+(\tau)_{\rm latt.}$ on a lattice of size 
$64^3\times16$ at $T=3T_c$ for several values of $\kappa$.
From the figure one sees that the shape of 
$S^{\rm M}_+(\tau)_{\rm latt.}$ approaches a single 
exponential function for small $\kappa$, while it 
becomes flat and symmetric as $\kappa$ becomes larger.
In the vicinity of the source, {\it i.e.} at small and 
large $\tau$, we see deviations from this generic picture,
which can be attributed to distortion effects arising 
from the presence of the source.
In fact, by comparing the correlation functions 
on the lattices with $N_\tau=12$ and $16$, 
we find that such a distortion is clearly seen only 
at the $\tau$ value closest to the source.
It is thus expected that this deviation arises only 
in the vicinity of the source and hence is negligible 
in the continuum limit.

In order to get control over distortion effects close to the source, 
we have performed fits using points 
$\tau_{\rm min} \le \tau \le N_\tau -\tau_{\rm min}$ with 
$\tau_{\rm min}=2,3,4$ and $5$ for $N_\tau=16$, and
$\tau_{\rm min}=2,3$ and $4$ for $N_\tau=12$.
We have checked that the dependence of the fitting 
parameters on $\tau_{\rm min}$ are small; 
the fit results obtained with different 
$\tau_{\rm min}$ coincide within the statistical error.
In the following analysis, we use $\tau_{\rm min}=4$ and 
$3$ for $N_\tau=16$ and $12$, respectively.

The resulting correlation functions in the two-pole ansatz, 
Eq.~(\ref{eq:2pole}), obtained from correlated fits
are shown in Fig.~\ref{fig:S} as solid lines.
One sees that $S_+^{\rm M}(\tau)_{\rm latt.}$ is 
well reproduced by our fitting ansatz. 
In Fig.~\ref{fig:chi2_p0}, we show the bare quark mass 
dependence of the $\chi^2/{\rm dof}$ on lattices of size 
$64^3\times16$ and $T/T_c=1.25, 1.5$ and $3$, where 
the critical hopping parameter $\kappa_c$ in Eq.~(\ref{eq:m_0}) 
will be defined in the next subsection.
The figure shows that $\chi^2/{\rm dof}$ is of order unity 
around $m_p=0$, which means that our fitting ansatz 
can describe the lattice correlation function well for light
quarks.
In particular, $\chi^2/{\rm dof}$ is less than unity
for $m_p/T \lesssim 0.3$ with $T/T_c=1.5$ and $3$.
For large bare quark masses and close to $T_c$, on the 
other hand, the two-pole ansatz eventually becomes worse.

The success of two-pole ansatz for the quark correlation 
functions indicates that the excitation modes of quarks 
near but above $T_c$ are good quasi-particles with small 
decay rates similar to those found in the perturbative region.
In terms of the complex pole of the propagator, the results 
suggest that 
the positions of the poles would be near the real axis 
at $\omega=E_1$ and $-E_2$ with small imaginary parts.
Provided that the positions of poles of the quark 
propagator are gauge independent \cite{g-dep}, 
this also suggests that our results on the
fitting parameters $E_1$ and $E_2$ have small gauge 
dependence.

\subsection{Pole structure}
\label{sec:pole_p0}

\begin{figure*}[tbp]
\begin{center}
\includegraphics[width=.7\textwidth]{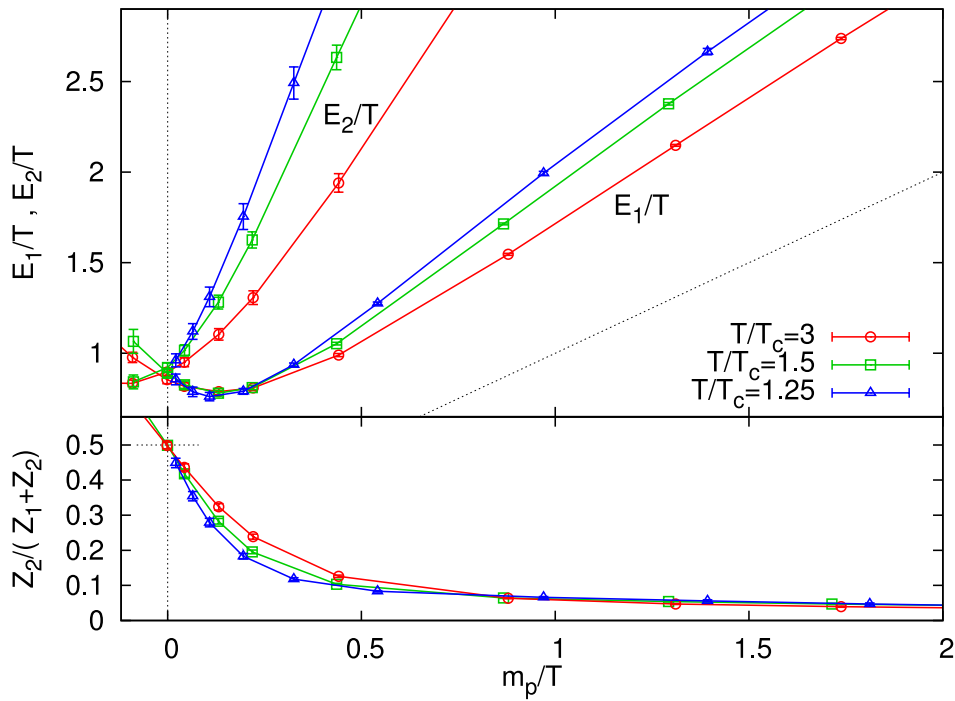}
\caption{
Bare quark mass dependence of fitting parameters $E_{1,2}$ 
and the relative strength of the plasmino mode, 
$Z_2 / ( Z_1+Z_2 )$, at $T/T_c=1.25$, $1.5$ and $3$ obtained 
from calculations on lattice of size $64^3\times16$.
}
\label{fig:ZE}
\end{center}
\end{figure*}

In Fig.~\ref{fig:ZE}, we show the dependence of 
$E_1$, $E_2$ and $Z_2 / ( Z_1+Z_2 )$ on the bare quark
mass for $T/T_c=1.25$, $1.5$, and $3$
obtained from calculations on 
lattices of size $64^3\times16$.
Error-bars have been estimated from a  Jackknife analysis.
The bare quark mass $m_p$ is defined in Eq.~(\ref{eq:m_p}) 
with $\kappa_c$ determined by the $\kappa$ dependence 
of $\rho^{\rm M}_+(\omega)$ as described below.

The figure shows that the ratio $Z_2 / (Z_1+Z_2)$ 
becomes larger with decreasing $m_p$ and eventually 
reaches $0.5$ irrespective of $T$.
The numerical result for each $T$ shows that $E_1=E_2$ 
is satisfied within statistical errors at this point.
These results show two important features of 
the structure of $\rho^{\rm M}_+(\omega)$ at this point.
First, since $\rho^{\rm M}_+(\omega)$ becomes an 
even function, the quark propagator is chirally 
symmetric at this point within the statistical error 
(see Sec.~\ref{sec:Dirac}).
Due to this feature, it is natural to define the hopping 
parameter satisfying $Z_1=Z_2=0.5$ to be the critical 
hopping parameter, $\kappa_c$. 
The values of $\kappa_c$ defined in this way 
is given in the second column of 
Table~\ref{table:p0_result}
\footnote{
Clearly, our definition of $\kappa_c$ introduced 
above is not unique.
Possible alternative definitions are, for example, 
the value of $\kappa$ at which (1) $E_1=E_2$, or at 
which (2) the correlation function in the scalar 
channel $S_{\rm s}(\tau,\bm{0})$ becomes smallest.
We have checked that the systematic error on $\kappa_c$ 
arising from these different definitions is of the same 
order as the statistical error on $\kappa_c$ given in 
Table~\ref{table:p0_result}.
}.
We have checked that these values are consistent 
with those obtained in \cite{Bielefeld,Datta:2003ww} 
from the vanishing of the isovector axial current.
The second observation is that 
$\rho^{\rm M}_+(\omega)$ at $\kappa=\kappa_c$ 
has the same form as the spectral function in the high 
temperature limit, Eq.~(\ref{eq:rho^M_HTL}).
We therefore define the thermal mass of the quark on the 
lattice as $m_T \equiv ( E_1+E_2 )/2$ at $\kappa=\kappa_c$.
The value of $m_T$ for each $T$ with $N_\sigma=64$
is given in the third column of 
Table~\ref{table:p0_result}.
One finds that the ratio $m_T/T$ is insensitive 
to $T$ in the range analyzed in this work, 
although it becomes slightly larger with decreasing $T$,
which would be in accordance with the expected parametric
form at high temperature, $m_T\sim gT$.

\begin{table}
\begin{center}
\begin{tabular}{cc|cc}
\hline
\hline
& & \multicolumn{2}{c}{$m_T/T$} \\
$T/T_c$ & $\kappa_c$     & $N_\sigma=64$ & $N_\sigma=\infty$ \\
\hline
$3$     & $0.133997(13)$ & $0.875(8)$  & $0.771(20)$ \\ 
$1.5$   & $0.134999(10)$ & $0.906(8)$  & $0.800(18)$ \\ 
$1.25$  & $0.135248(10)$ & $0.899(12)$ & $0.803(24)$ \\ 
\hline
\hline
\end{tabular}
\end{center}
\caption{
The second column shows the critical hopping 
parameter $\kappa_c$ determined from $\kappa$ 
dependence of the fitting functions for lattices 
of size $64^3\times16$.
The values of the thermal mass $m_T$ analyzed 
on lattice with $N_\sigma=64$, and that extrapolated 
to the infinite volume limit $N_\sigma=\infty$ 
are also presented in the right columns.
}
\label{table:p0_result}
\end{table}

Figure~\ref{fig:ZE} also shows that the relative strength 
of the plasmino pole,
$Z_2/(Z_1+Z_2)$, decreases with increasing values
of the bare mass, $m_p$. The spectral function
$\rho^{\rm M}_+(\omega)$ thus will
eventually be dominated by a single-pole.
This result agrees with the generic observations discussed in 
Sec.~\ref{sec:rho:special}, {\it i.e.}  $\rho^{\rm M}_+(\omega)$ 
approaches the spectral function of free quarks, Eq.~(\ref{eq:rho_Yukawa2}),
as the bare quarks mass becomes larger
(see also Appendix~\ref{sec:Yukawa}).
The quark mass dependence of the fitting parameters at large 
$m_p$ thus is reasonable.
We also note that $E_1$ has a minimum at $m_p>0$,
while $E_2$ is a monotonically increasing function.
This is in contrast to
the one-loop result in the Yukawa model, 
summarized in Appendix~\ref{sec:Yukawa}, where 
the position of the peak in $\rho^{\rm M}_+(\omega)$ 
at positive energy corresponding to $E_1$ is 
a monotonically increasing function of $m/T$, 
while the absolute value of that at negative energy, $E_2$,
decreases monotonically.
The $m_p$ dependence of $E_1$ and $E_2$ determined from
our lattice-QCD calculations
therefore is qualitatively different from the 
perturbative result.
The non-perturbative nature of the gluon field could be 
responsible for this behavior.
Indeed, the minimum of $E_1$ becomes shallower with increasing
temperature and the slope of $E_2$ decreases, as can be seen in
Fig.~\ref{fig:ZE}.  The perturbative behavior thus may well 
be recovered in the perturbative high temperature limit.

\begin{figure}[tbp]
\begin{center}
\includegraphics[width=.49\textwidth]{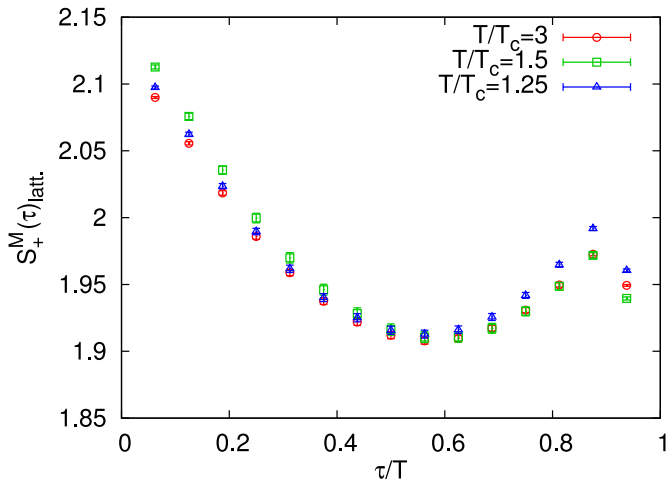}
\caption{
Lattice correlation functions near the chiral limit
for $T/T_c=1.25$, $1.5$ and $3$.
The value of $m_p$ is $m_p/T \simeq 0.1$, $0.05$ and $0.08$ 
for each $T$, respectively.
}
\label{fig:corr6416T}
\end{center}
\end{figure}

Finally, we shall briefly discuss the $T$ dependence of 
the magnitude of the residues $Z_1$ and $Z_2$. 
In Fig.~\ref{fig:corr6416T} we show 
the correlation function $S^{\rm M}_+(\tau)$ near the chiral 
limit, $m_p/T\simeq 0.08$, for $T/T_c=1.25$, $1.5$ and $3$.
The figure shows that the magnitude of $S^{\rm M}_+(\tau)$ 
is insensitive to $T$.
This result indicates that the magnitude of both residues, 
$Z_1$ and $Z_2$, does not have a strong $T$ dependence
for $T/T_c\gtrsim1.25$.

\subsection{Beyond the chiral limit}
\label{sec:m_p<0}

\begin{figure}[tbp]
\begin{center}
\includegraphics[width=.49\textwidth]{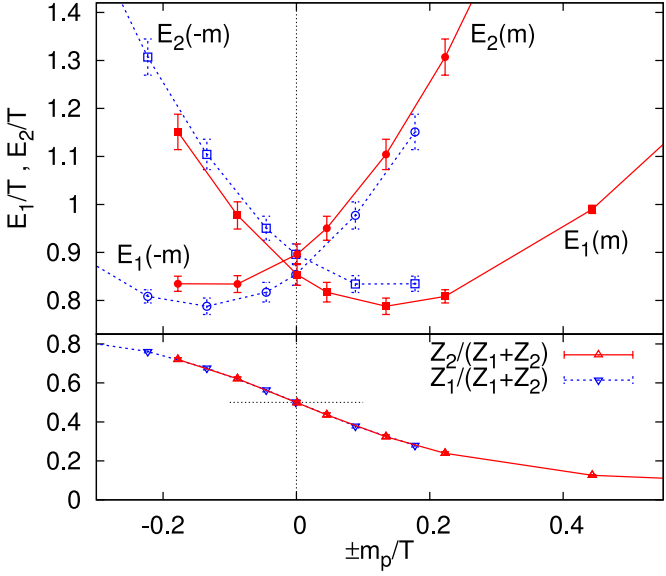}
\caption{
Fitting parameters $E_{1,2}$ and $Z_2/(Z_1+Z_2)$
near the chiral limit for $T/T_c=3$.
The dotted lines are $E_{1,2}$ as functions of $-m_p$,
and $Z_1/(Z_1+Z_2)$.
}
\label{fig:m0}
\end{center}
\end{figure}

On the lattices above $T_c$, one can solve 
Eq.~(\ref{eq:psiKpsi}) in and even beyond the chiral 
limit, since chiral symmetry is not spontaneously 
broken above $T_c$ and the numerical calculation does 
not suffer from the light Nambu-Goldstone mode.
From Eq.~(\ref{eq:m_p}), the hopping parameter for 
$\kappa>\kappa_c$ corresponds to the negative Dirac mass.
In Fig.~\ref{fig:m0}, we show $m_p$ dependence of the
fitting parameters near the chiral limit for $T/T_c=3$
\footnote{ 
We have checked that correlators other 
than those in the vector and scalar channels vanish within
statistical errors even for $\kappa>\kappa_c$.
}.
We plot the numerical results only for 
$m_p\gtrsim-0.2$, 
since the convergence of the inversion routine to solve 
Eq.~(\ref{eq:psiKpsi}) based on the BiCGStab algorithm 
starts failing there.

If the system possesses a charge conjugation symmetry, 
the sign of the Dirac mass does not affect any observables.
One can, however, show that the roles of 
$\rho^{\rm M}_+(\omega)$ and $\rho^{\rm M}_-(\omega)$ are 
exchanged when the sign of the Dirac mass is reversed;
\begin{align}
\rho^{\rm M}_+(\omega)|_{m_p=-m}
=\rho^{\rm M}_-(\omega)|_{m_p=m}.
\end{align}
This formula is shown by the fact that 
$\rho_0(\omega,p)$ and $\rho_{\rm s}(\omega,p)$ are even 
and odd, respectively, as functions of the bare quark mass, 
and Eqs.~(\ref{eq:rho_02}) and (\ref{eq:rho_s2}).
In terms of the fitting parameters in 
the two-pole ansatz Eq.~(\ref{eq:2pole}),
this requires that 
\begin{align}
E_1(\pm m)=E_2(\mp m) \mbox{ , and  }
Z_1(\pm m) = Z_2(\mp m).
\label{eq:-m}
\end{align}
In Fig.~\ref{fig:m0}, $E_1$ and $E_2$ as functions of 
$-m_p$ and $Z_1/(Z_1+Z_2)$ are shown by the dotted lines.
One sees from the figure that Eq.~(\ref{eq:-m})
is approximately satisfied within the statistical error.
This result shows that our numerical result 
behaves reasonably around the chiral limit.
The similar result is obtained on lattices for 
$T=1.5T_c$.

\subsection{Volume and lattice spacing dependence}
\label{sec:volume}

\begin{figure}[tbp]
\begin{center}
\includegraphics[width=.49\textwidth]{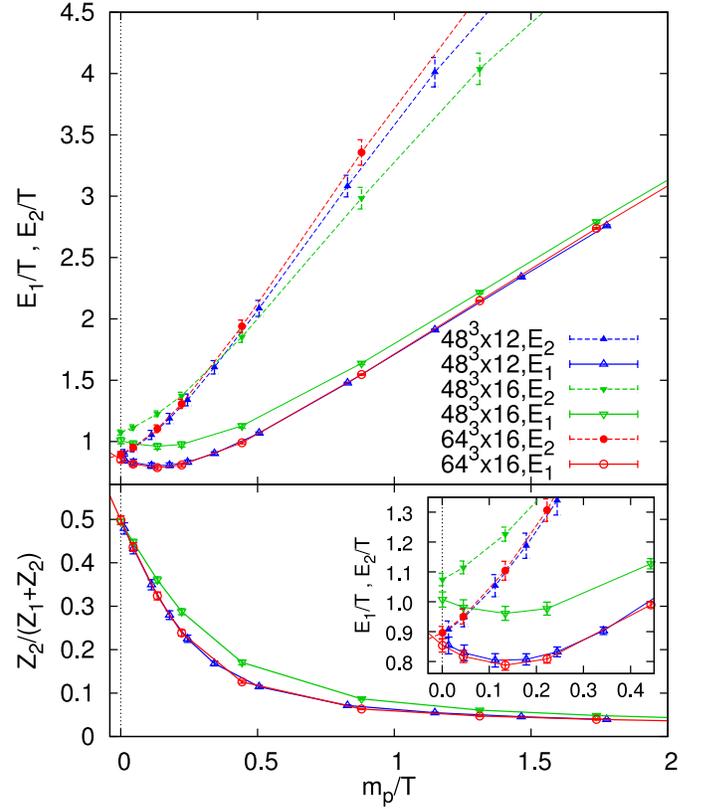}
\caption{
Bare quark mass dependence of parameters $E_1$, $E_2$,
and $Z_2/(Z_1+Z_2)$ at $T=3T_c$ for lattices of size
$64^3\times16$, $48^3\times16$ and $48^3\times12$.
}
\label{fig:compare}
\end{center}
\end{figure}

In order to check the dependences of our results 
on the lattice spacing and finite volume, we analyzed 
the quark propagator at $T/T_c=3$, $1.5$ and $1.25$ 
on lattices with different $a$ and $N_\sigma$ 
as shown in Table~\ref{table:param}. 
Results for $E_1$, $E_2$ and $Z_2/(Z_1+Z_2)$ 
obtained at $T/T_c=3$ for two different values of 
the lattice cut-off and two different physical 
volumina are shown in Fig.~\ref{fig:compare}. 
Comparing the results obtained on lattices with 
different lattice cut-off, $a$, but same physical 
volume, {\it i.e.} $64^3\times16$ and $48^3\times12$, 
one sees that any possible cut-off dependence 
is statistically not significant in our analysis. 
On the other hand we find a clear dependence of 
these quantities on the spatial volume; 
when comparing lattices with aspect ratios 
$N_\sigma/N_\tau =3$ and $4$ we find that the 
energy levels, $E_1$ and $E_2$, drop significantly
near the chiral limit as the volume is increased.
For larger values of the bare mass $m_p$  the volume 
dependence of $E_1$ becomes small.
A similar behavior is observed 
for $T/T_c=1.5$ and $1.25$.

\begin{figure}[tbp]
\begin{center}
\includegraphics[width=.49\textwidth]{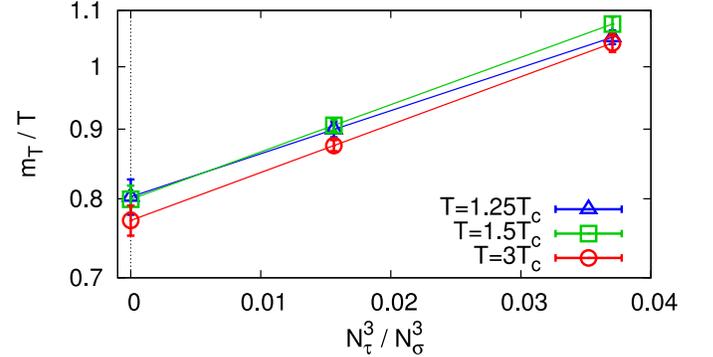}
\caption{
Extrapolation of the thermal mass of the quark
to the infinite volume limit for $T/T_c=3$, $1.5$, 
and $1.25$.
}
\label{fig:mt_extrpl}
\end{center}
\end{figure}

The presence of a strong volume dependence of the 
quark propagator is not unexpected. 
In fact, for the emergence of the thermal mass
hard gluons having momenta $p\simeq T$ play 
a crucial role \cite{HTL,LeBellac}. 
However, the lowest non-vanishing gluon momentum on the 
lattice is, $p_{\rm min}/T = 2\pi (N_\tau / N_\sigma)$, 
which still is larger than unity on lattices 
with aspect ratio $N_\sigma/N_\tau=4$. 
In fact, at present one cannot
rule out that the
discretization of low momenta can also be 
responsible for the success of the two-pole ansatz 
for $\rho_+^{\rm M}(\omega)$, since scattering 
processes giving rise to the width of quasi-particles 
can be suppressed due to the discretization of momentum.
An analysis of quark spectral functions on lattices
with even larger spatial volume, or other than periodic
boundary conditions, is needed in the 
future to properly control effects of small momenta. 

In the present study, we estimate the thermal mass $m_T$
in the $V\to\infty$ limit by extrapolating the results 
obtained for two different volumina with $N_\tau=16$.
The extrapolation of $m_T$ with the ansatz for the volume
dependence 
$m_T(N_\tau/N_\sigma)= m_T(0) \exp(c N_\tau^3/N_\sigma^3)$
for each temperatures is shown in Fig.~\ref{fig:mt_extrpl}.
This extrapolation suggests that 
finite volume effects may still be of the order of $15\%$ 
in our current analysis of $m_T/T$. 
The value of $m_T(0)$ determined from this extrapolation 
is depicted in the far right columns of 
Table.~\ref{table:p0_result}.

\subsection{Quark mass dependence}

\begin{table}
\begin{center}
\begin{tabular}{c|cc|cc}
\hline
\hline
$T/T_c$ & $\kappa$  & $\kappa_c$ & $m_D$[GeV] & $Z_2/(Z_1+Z_2)$ \\
\hline
$3$     & $0.13114$ & $0.13454(3)$ & $1.625(5)$ & $0.057(2)$ \\
$1.5$   & $0.1290$  & $0.13540(3)$ & $1.534(6)$ & $0.042(2)$ \\
\hline
\hline
\end{tabular}
\end{center}
\caption{
Pole mass of the charm quark $m_D=E_1$ and the strength
of the plasmino mode $Z_2/(Z_1+Z_2)$ on $48^3\times12$ lattice
for $T/T_c=3$ and $1.5$.
The parameter $\kappa$ for the charm quark is 
those employed in \cite{Datta:2003ww}.
}
\label{table:charm}
\end{table}

Here we want to discuss quasi-particle properties of the 
physical quarks, {\it i.e.} up, down and charm. 
So far we have treated the bare quark mass 
as a free parameter thus. Clearly one can discuss the 
properties at physical values of the quark masses by 
choosing the bare mass $m_p$ appropriately.
Such information may be exploited for the understanding 
of the QGP phase near $T_c$.

\begin{figure*}[tbp]
\begin{center}
\includegraphics[width=.7\textwidth]{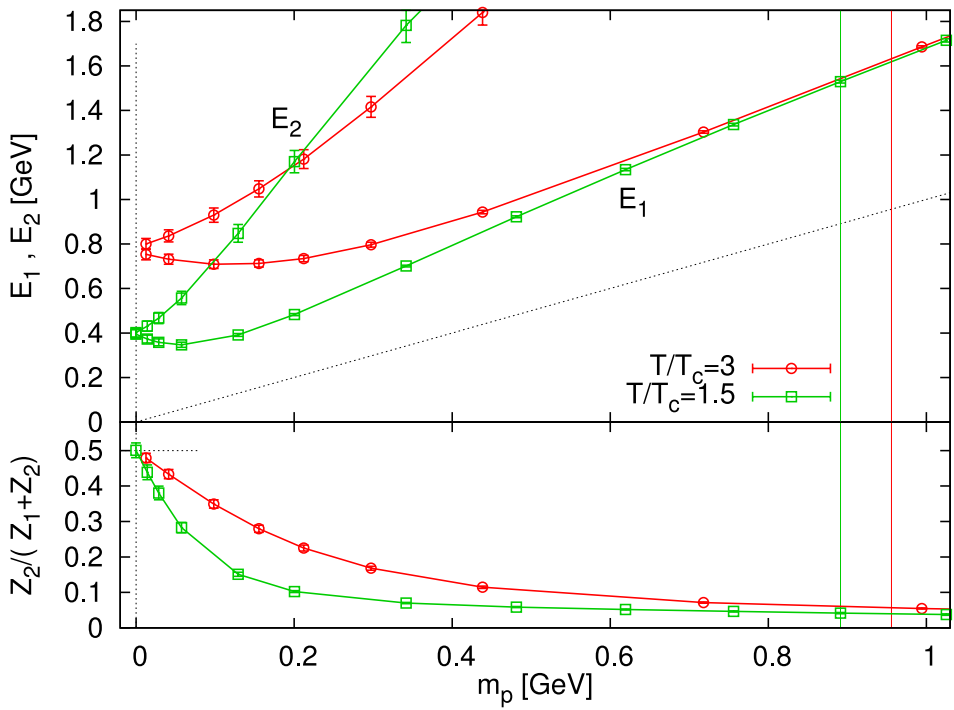}
\caption{
Dependences of $E_1$, $E_2$ and $Z_2/(Z_1+Z_2)$ on
bare quark mass $m_p$ on $48^3\times12$ lattice
for $T/T_c=3$ and $1.5$ in the physical unit.
The value of $m_p$ corresponding to the charm quark
employed in \cite{Datta:2003ww} are shown by the vertical lines.
}
\label{fig:pole_GeV}
\end{center}
\end{figure*}

In order to discuss properties of the quark propagator 
for physical quark mass values,
we first show the bare quark mass, $m_p$, dependences
of the fitting parameters $E_1$, $E_2$ and $Z_2/(Z_1+Z_2)$
in physical units (GeV) in Fig.~\ref{fig:pole_GeV}.
Throughout this subsection, we use lattices of size 
$48^3\times12$ for $T/T_c=1.5$ and $3$;
these sets of gauge configurations are exactly those used
also in the analysis of charmonia at finite $T$ in 
\cite{Datta:2003ww}. They are therefore most suitable for the comparison
between properties of quarks and charmonia analyzed there.
As discussed before, the lattice spacing dependences of 
these results are small and the figure hardly changes even 
if we employ the finer lattices of size $64^3\times16$.

Let us first investigate the quasi-particle property 
of the charm quark.
For this purpose we can use the values of $\kappa$ 
corresponding to the charm quark evaluated in 
\cite{Datta:2003ww}, which are shown in Table~\ref{table:charm}.
In Fig.~\ref{fig:pole_GeV}, the value of $m_p$ corresponding 
to these $\kappa$ values is shown for each $T$  
by a vertical line.
One sees that the values of $Z_2/(Z_1+Z_2)$ on these 
lines, which are shown in the far right column of 
Table~\ref{table:charm}, are small, $Z_2/(Z_1+Z_2)\ll 1$.
This means that the strength of the plasmino mode 
is weak and the structure of the quark spectral function 
is close to that of free quarks, 
Eq.~(\ref{eq:rho^M_free}), with a single pole at $\omega=E_1$.
Therefore, it should be reasonable to regard the charm 
quarks at these temperatures as free Dirac particles 
with a Dirac mass $m_D=E_1$.
The value of $m_D$ for each $T$ is shown 
in the fourth column of Table~\ref{table:charm}.

The lattice simulations suggest the existence of 
sharp peaks in the spectral function of $\eta_c$ 
and $J/\psi$ even above $T_c$ up to $T=1.5-2T_c$ 
\cite{charm,Datta:2003ww}.
It is interesting to compare the Dirac mass of the 
charm quark obtained here with the spectral 
functions of charmonia.
In particular, twice the Dirac mass, $2m_D$, gives 
a threshold for the decay process of the charmonia, 
provided that the potential between a quark and an 
antiquark vanishes at long distances.
The numerical result in \cite{Datta:2003ww} 
shows that the energies of the peaks corresponding to 
$\eta_c$ and $J/\psi$ for $T/T_c=1.5$ are 
$m_{\eta_c}\simeq3.4$GeV and $m_{J/\psi}\simeq3.8$GeV
\footnote{
We note that the parameters used to determine the physical
scale used in \cite{Datta:2003ww} and the present study 
are slightly different.
The masses in physical unit quoted here take
this difference into account for comparison, {\it i.e.}
we use our value for $T_c$ to set the scale.}.
These values are clearly larger than $2m_D\simeq3.1$GeV.

If the confinement potential is screened completely,
$m_{\eta_c}$ and $m_{J/\psi}$ thus are resonance 
states inside the continuum. At least at 
$T/T_c\lesssim 1.5$
the heavy quark free energy still has a complicated
structure which still shows a linear rise over the 
distance range relevant for quarkonium physics
\cite{Kaczmarek:2004gv}. This needs to be taken into
account in any further quantitative discussion.

Here, we note that the values of $\kappa$ 
employed in \cite{Datta:2003ww} are not accurately 
corresponding to the physical charm quark:
With these parameters 
the masses corresponding to $\eta_c$ in the vacuum
are about $3.4$GeV and $4.1$GeV on each lattice 
for $T/T_c=1.5$ and $3$, respectively.  They are therefore slightly 
larger than the experimental value.
These parameters therefore should be interpreted as 
a guide for the charm quark.
In particular, the values of $m_D$ in 
Table~\ref{table:charm} are not the exact values 
for the charm quark.
It should, however, be emphasized that the above arguments
about the comparison between $2m_D$ and masses of 
charmonia makes sense, because the same hopping parameter
is employed in both analyses.

Finally, we turn to a discussion on the light quarks.
Figure~\ref{fig:pole_GeV} shows that the conditions 
$Z_2/(Z_1+Z_2)\simeq0.5$ and $E_1\simeq E_2$ are 
satisfied at $m_p$ corresponding to the light quark masses, 
say $m_q\simeq0.01$GeV, for each temperature.
This shows that the effect of a non-zero $m_p$ is negligible 
and the quasi-particle picture for light quarks is close 
to that obtained in the high-temperature limit, 
Eq.~(\ref{eq:rho^M_HTL}), 
{\it i.e.}
light quarks have a thermal mass $m_T$.
This quasi-particle property of the light quarks 
suggests that the effect of the thermal mass of 
light quarks should be taken into account when one 
consider quark quasi-particles as the basic ingredients
in the studies of thermodynamics \cite{EoS},
quarkonia and the chiral transition \cite{Hidaka:2006gd}
of the QGP phase near $T_c$.
The value of the bare mass for the strange quarks, 
$m_q\simeq0.08$~GeV, on the other hand, is in the 
intermediate region between the two simple limits
for these temperatures.

\section{Quark propagator at finite momentum}
\label{sec:p}

In this section, we analyze the quark spectral function 
at finite momentum on lattices with size $64^3\times16$
for $T/T_c=1.5$ and $3$.
Throughout this section we consider the quark propagator 
in the chiral limit by fixing $\kappa=\kappa_c$, 
where $\kappa_c$ is the critical hopping parameter 
determined in the previous section.
The correlation function on the lattice is calculated using 
a wall source, Eq.~(\ref{eq:wall}), with momentum $\bm{p}$.
The quark propagator in the chiral limit is 
decomposed into $\rho^{\rm P}_\pm(\omega,p)$
according to Eq.~(\ref{eq:rho^P_pm}).
Following the same approach used in Sec.~\ref{sec:T>Tc}
at zero momentum, we adopt the two-pole ansatz 
\begin{align}
\rho_+^{\rm P}(\omega,p)
= Z_1 \delta( \omega - E_1 ) + Z_2 \delta( \omega + E_2 ) ,
\label{eq:2pole_p}
\end{align}
and determine four parameters from a correlated fit
with $\tau_{\rm min}=4$.
The $\delta$-functions at $\omega=E_1$ and $-E_2$ correspond
to the normal and plasmino modes, respectively.
We found that $\chi^2/{\rm dof}$ with this ansatz is always 
smaller than $1.5$ for all momenta analyzed in this study.
This result means that the two-pole ansatz again reproduces 
the lattice correlation function well.

\begin{figure}[tbp]
\begin{center}
\includegraphics[width=.49\textwidth]{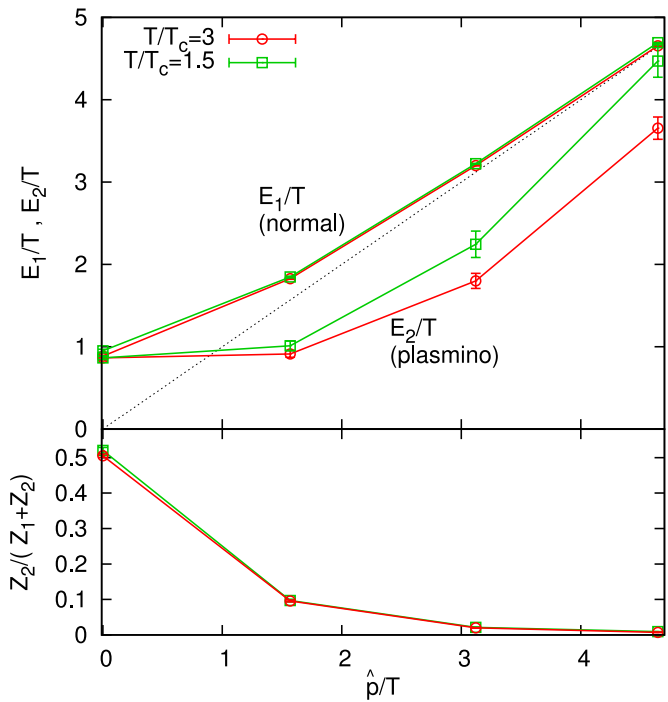}
\caption{
Dependences of the fitting parameters 
$E_1$ and $E_2$ and the ratio $Z_2/(Z_1+Z_2)$ on
the lattice momentum $\hat{p}=(1/a) \sin(pa)$
for $T/T_c=1.5$ and $3$.
}
\label{fig:p>0}
\end{center}
\end{figure}

In Fig.~\ref{fig:p>0}, we show the momentum dependence
of the fitting parameters $E_1$, $E_2$, and 
$Z_2/(Z_1+Z_2)$ for $T/T_c=1.5$ and $3$.
The horizontal axis represents the momentum 
on the lattice $\hat{p}=(1/a) \sin pa$.
The figure shows that for large momentum 
$Z_2/(Z_1+Z_2)$ rapidly decreases and $E_1$ approaches 
the light cone.
The spectral function at large momentum therefore 
approaches that of a free quark, consistent with
the perturbative result.
One also finds that $E_2$ is always smaller than $E_1$,
in contrast to the results in Sec.~\ref{sec:T>Tc}.
This behavior qualitatively agrees with the behavior of
poles in the high $T$ limit \cite{LeBellac}.
One also observes from Fig.~\ref{fig:p>0} that $E_2$ enters
the space-like region at high momentum.
While in one-loop perturbation theory the plasmino mode
always exists in the time-like region, higher order 
corrections could give rise to such behavior of 
the plasmino mode.
At least, such behavior does not contradict causality.

An interesting property of the quark propagator in the 
high temperature limit Eq.~(\ref{eq:S_HTL}) is that the 
dispersion relation
of the plasmino has a minimum at finite momentum.
In Fig.~\ref{fig:p>0}, one sees that the value of $E_2$ 
at lowest non-zero momentum on our lattice, 
$ p_{\rm min} = 2\pi T(N_\tau/N_\sigma) \simeq 1.5T $,
is slightly larger than that at zero momentum,
and the existence of such a minimum is 
suggested but not yet confirmed in the present analysis.
A more detailed analysis at smaller momenta 
would clearly be needed, which requires calculations
on lattices with a larger aspect ratio $N_\sigma/N_\tau$.

The quark spectral function at high temperatures, 
Eq.~(\ref{eq:rho^P_HTL}), has a continuum 
$\rho_{\rm cont.}(\omega,p)$ in the space-like region, 
which physically originates from the Landau damping.
At leading order, the spectral weight of 
$\rho_{\rm cont.}(\omega,p)$, 
$\int_{-p}^{p} d\omega \rho_{\rm cont.}(\omega,p)$, 
becomes $0.2$ at most.
The success of the two-pole fit for 
$\rho^{\rm P}_+(\omega,p)$ without the continuum 
therefore seems inconsistent with the perturbative result.
A possible reason for this feature is the discretization 
of momenta on the lattice, since the Landau damping giving 
rise to $\rho_{\rm cont.}(\omega,p)$ can be suppressed 
due to the missing momenta $p\lesssim T$ in our 
current analysis.
Lattices with much larger spatial volume are required to 
clarify this problem as well as the detailed properties 
of the dispersion relations including the minimum of the 
plasmino mode.

\begin{figure}[tbp]
\begin{center}
\includegraphics[width=.49\textwidth]{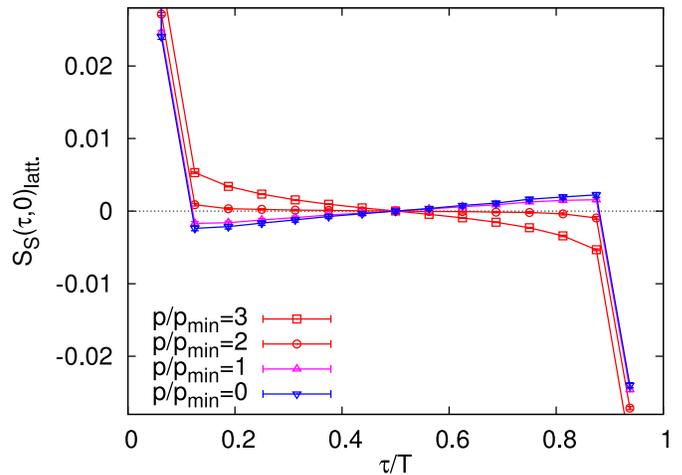}
\caption{
Scalar channel of lattice correlation function 
$S_{\rm s}(\tau,0)_{\rm latt.}$ at $\kappa=\kappa_c$
for $T/T_c=1.5$ with several values of $p$.
}
\label{fig:fm_m}
\end{center}
\end{figure}

So far, we discussed the spectral functions 
$\rho^{\rm P}_\pm(\omega,p)$, assuming that the scalar 
channel $\rho_{\rm s}(\tau,p)$ vanishes.
In order to check the validity of this assumption,
we show in Fig.~\ref{fig:fm_m} the momentum dependence 
of the correlation function in the scalar channel, 
$S_{\rm s}(\tau,p)_{\rm latt.}$, for $T/T_c=1.5$.
The figure shows that the absolute values of 
$S_{\rm s}(\tau,p)_{\rm latt.}$ are smaller than $0.005$ 
up to $p/p_{\rm min}\simeq3$ except for $\tau$-values 
next to the source where they suffer from lattice artifacts.
These values are more than two orders smaller than 
the typical values of $S^{\rm P}_+(\tau,p)_{\rm latt.}$,
and thereby being negligibly small, indeed.
The figure shows that deviations of 
$S_{\rm s}(\tau,p)_{\rm latt.}$ from zero
become statistically significant as $p$
increases. This is a lattice artefact and is expected 
to arise from the momentum dependence of the 
mass term in the Wilson formulation; for 
free Wilson fermions the mass term is given by
$M(p) = m_0 + r(1 - \cos pa)$
with $r$ being the Wilson parameter.
The fact that $S_{\rm s}(\tau,p)$ is still small
even at $p=3p_{\rm min}\simeq 4.5T$  shows
that our lattice is fine enough so that the effect of
explicit chiral symmetry breaking, which arises
from the Wilson term, is well suppressed.

\section{Quark propagator below $T_c$}
\label{sec:T<Tc}

\begin{figure}[tbp]
\begin{center}
\includegraphics[width=.49\textwidth]{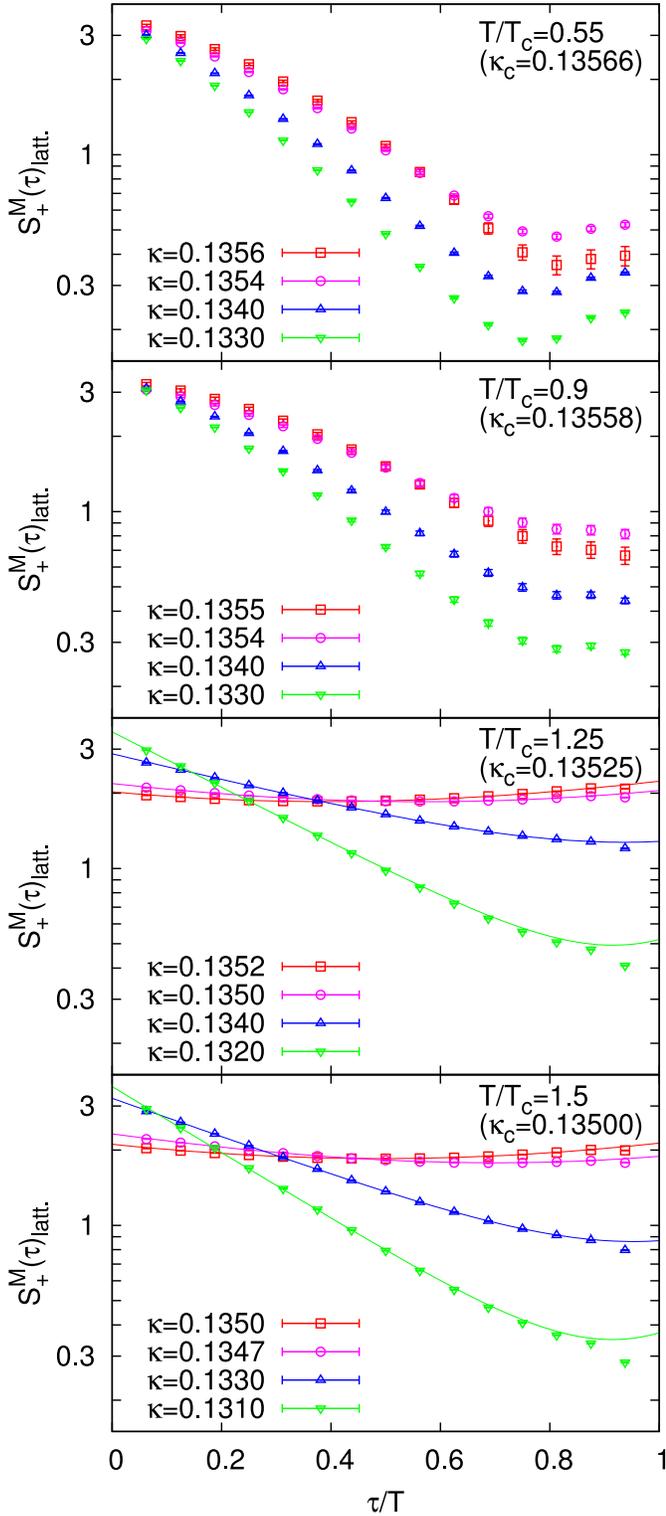}
\caption{
Lattice correlation functions $S^{\rm M}_+(\tau)_{\rm latt.}$
for several values of $T$ below and above $T_c$ 
on $48^3\times16$ lattices.
In each panel, $S^{\rm M}_+(\tau)_{\rm latt.}$ is depicted
for four values of $\kappa$; two of which are close to $\kappa_c$.
In the lower two panels for $T>T_c$, the correlation functions 
obtained by the two-pole ansatz is also shown by solid lines.
}
\label{fig:below}
\end{center}
\end{figure}

Next, we analyze the quark correlation function below $T_c$.
In this section, we restrict the analysis 
to the lattices of size $48^3\times16$ for simplicity.
In the upper two panels of Fig.~\ref{fig:below}, 
we show the correlation function at zero momentum,
$S^{\rm M}_+(\tau)_{\rm latt.}$, 
for $T/T_c=0.5$ and $0.93$ and for
several values of $\kappa$.
The critical hopping parameter $\kappa_c$ for each $T$ 
determined in \cite{Bielefeld} is $0.13566$ and
$0.13558$, respectively.
For comparison, $S^{\rm M}_+(\tau)_{\rm latt.}$ above $T_c$
for $T/T_c=1.25$ and $1.5$ is shown in the lower two panels.

Before starting the discussion of results obtained below 
$T_c$, we first recapitulate the qualitative behavior 
of $S^{\rm M}_+(\tau)_{\rm latt.}$ above $T_c$.
As we have seen in Sec.~\ref{sec:T>Tc}, the following two 
qualitative features are observed above $T_c$:
(1) 
$S^{\rm M}_+(\tau)_{\rm latt.}$ is well reproduced by
the two-pole ansatz Eq.~(\ref{eq:2pole}). The fitting 
results are shown by solid lines in the lower two panels.
(2)
$S^{\rm M}_+(\tau)_{\rm latt.}$ 
approaches a symmetric function, Eq.~(\ref{eq:S_chiral}), 
as $\kappa\to\kappa_c$, which means that chiral 
symmetry of the quark propagator is recovered there.

The upper two panels in Fig.~\ref{fig:below} clearly show 
that the behavior of $S^{\rm M}_+(\tau)_{\rm latt.}$ 
below $T_c$ is qualitatively different from those above $T_c$.
First, $S^{\rm M}_+(\tau)_{\rm latt.}$ is concave
in the log-scale plot at $\tau/T\lesssim0.6$ for any value of $\kappa$.
Such structure can never be reproduced by the two-pole 
ansatz Eq.~(\ref{eq:2pole}).
In fact, we have checked that the fits with 
Eq.~(\ref{eq:2pole}) always gives unacceptably large 
$\chi^2/{\rm dof}$ below $T_c$.
Moreover, this behavior of $S^{\rm M}_+(\tau)_{\rm latt.}$ 
cannot be reproduced even if we use more than three
poles with positive residues.
Our result thus indicates the violation of positivity of 
$\rho^{\rm M}_+(\omega)$ below $T_c$, which is found 
also in the Schwinger-Dyson approach below $T_c$ \cite{SDE}.

The failure of the two-pole ansatz for $\rho^{\rm M}_+(\omega)$ 
indicates the absence of quasi-particles corresponding to
sharp peaks in $\rho^{\rm M}_+(\omega)$,
and this result seems consistent with a na\"ive picture
that quark excitations are confined below $T_c$.

In the last sections, we discussed that the gauge dependence
of our result is expected to be small, due to the success of
the two-pole approximation and the argument that the position 
of poles of propagators is gauge independent.
This argument breaks down below $T_c$, since we no longer 
conclude anything about the position of poles.
The violation of positivity of spectral functions could 
be an artifact of a specific choice of gauge fixing condition
\cite{BBS92}.
The calculation of $S^{\rm M}_+(\tau)_{\rm latt.}$ 
with different gauge fixing conditions may provide us 
with further clues to understand this problem.

\begin{figure}[tbp]
\begin{center}
\includegraphics[width=.49\textwidth]{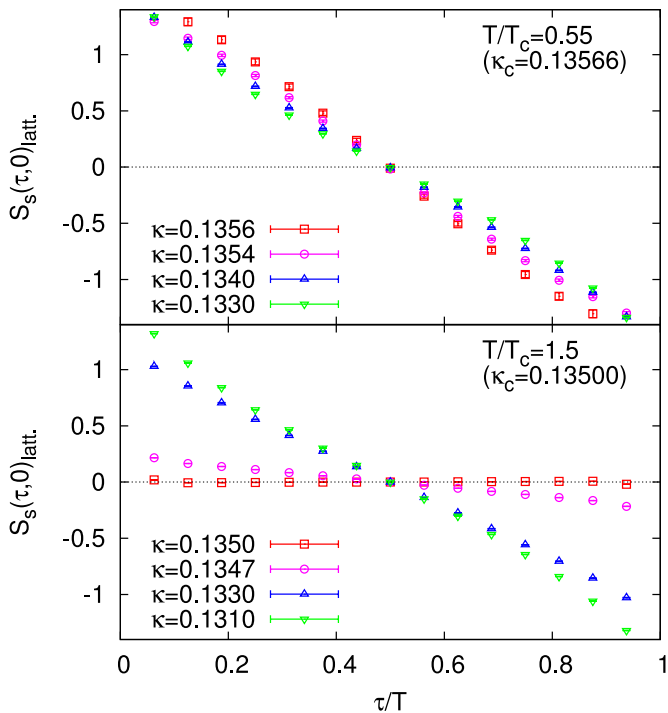}
\caption{
Scalar channel of lattice correlation function 
$S_{\rm s}(\tau,0)_{\rm latt.}$
for $T/T_c=0.5$ and $1.5$.
}
\label{fig:S_below}
\end{center}
\end{figure}

From Fig.~\ref{fig:below}, one also finds that 
$S^{\rm M}_+(\tau)_{\rm latt.}$ below $T_c$ does not 
approach a symmetric function as $\kappa\to\kappa_c$ 
in contrast to those above $T_c$.
This means that the quark propagator does not become chirally 
symmetric even in the chiral limit, which is consistent 
with spontaneous chiral symmetry breaking below $T_c$.
To see this behavior more clearly, we show in 
Fig.~\ref{fig:S_below} the correlation function in the 
scalar channel, $S_{\rm s}(\tau,0)_{\rm latt.}$, for 
several values of $\kappa$, for $T=0.5T_c$ and $1.5T_c$.
The figure shows that $S_{\rm s}(\tau,0)_{\rm latt.}$ 
for $T=0.5T_c$ indeed stays finite in the limit
$\kappa\to\kappa_c$, while that for $T=1.5T_c$ 
is vanishingly small in this limit.

Since the difference between the correlation functions 
below and above $T_c$ is quite remarkable, we conclude 
that the thermal modification of gluon fields during 
the deconfinement transition strongly affects also
the excitation properties of quarks propagating in 
this background field.
Since our fit ansatz fails in the confined phase, 
however, the detailed structure of the quark propagator 
is less clear.
The comparison with the quark propagator at $T=0$
calculated in lattice simulations \cite{T=0} and 
the Schwinger-Dyson equation \cite{SDE} will give us 
insights to understand the present results.
Such study, however, is beyond the scope of 
the present work.

\section{Summary}
\label{sec:summary}

In this publication, we analyzed the dependence of the quark 
spectral function on temperature $T$, bare quark mass $m$, and 
momentum $p$ in quenched lattice QCD with Landau gauge fixing.
Above $T_c$, we found that the two-pole approximations for 
the spectral functions in the projected channels, 
$\rho^{\rm M}_\pm(\omega)$ and $\rho^{\rm P}_\pm(\omega,p)$, 
can well reproduce the lattice correlation functions.
Although further studies on the volume dependence is
needed, this result indicates that the excitations of
quarks have small decay width even near $T_c$.
Below $T_c$, on the other hand, the two-pole ansatz fails 
completely and the behavior of quark correlation functions 
indicates the violation of positivity of the spectral function.

By analyzing the $m$ and $p$ dependence of these poles,
we confirmed that above $T_c$ two poles, the normal and 
plasmino excitations, appear in the quark propagator. 
The mass gaps of these excitation spectra should be 
interpreted as thermal masses, rather than Dirac mass.
It is found that the ratio $m_T/T\simeq0.8$ is insensitive 
to $T$ in the range analyzed in this study.
As the bare quark mass is increased, the spectral 
function eventually changes its form from that having 
a thermal mass to the free quark form.
We found that the bare quark mass of the charm 
quark is close to the latter, having a single mode 
with a Dirac mass.

All analyses of the present study are based on
the quenched approximation.
Although this approximation includes the leading contribution
in the high temperature limit \cite{LeBellac} and 
thus is valid at sufficiently high $T$, 
the validity of this approximation near $T_c$ is nontrivial.
For example, screening of gluons due to the 
polarization of the vacuum with virtual quark antiquark pairs
is neglected in this approximation.
The coupling to possible mesonic excitations 
\cite{HK85,charm,Datta:2003ww}, which may cause interesting 
effects in the spectral properties of the quark 
\cite{KKN06,Kitazawa:2007ep}, are not incorporated, either.
The comparison of the quark propagator between quenched and
full lattice QCD simulations would tell us 
the importance of these effects near $T_c$.
It also would be interesting to calculate perturbatively higher 
order corrections to the quasi-particle
properties of quarks \cite{Harada:2008vk}.
Such a calculation will help to clarify the origin of 
differences in the mass and momentum dependence of
the quark propagator found in our lattice calculation
in comparison to leading order perturbation theory.

There are many open questions that need to be 
analyzed in more detail in future calculations.
A numerical simulation with a large spatial 
volume is an important subject among them.
As discussed in the text, the influence of momenta 
smaller than the temperature is not properly included in the
present simulations; the smallest non-zero momentum
possible on our lattices with aspect ratio 4 and periodic
boundary conditions is larger than $T$.
Lattices allowing for momenta less than $T$
will be necessary to systematically analyze the
importance of low momentum excitations.
The existence of a minimum in the plasmino mode
could also be confirmed in such a study.
The calculation of the quark propagator in a different 
gauge is also important. It will allow to check directly 
the gauge dependence of the present results.
These subjects will be studied elsewhere.

\section*{Acknowledgments}
\label{ackn}
The lattice simulations presented in this work have been 
carried out using the cluster computers ARMINIUS@Paderborn, 
BEN@ECT* and BAM@Bielefeld.
M.~K. is supported by a Grant-in-Aid for Scientific 
Research by Monbu-Kagakusyo of Japan (No. 19840037).
This work has been supported in part by contract 
DE-AC02-98CH10886 with the U.S. Department of Energy.

\appendix

\section{Quark spectral function in Yukawa model}
\label{sec:Yukawa}

In this appendix, we review the spectral function of 
massive fermions coupled to a massless scalar boson 
via the Yukawa coupling at finite temperature $T$.
While the results given in this appendix are essentially 
the same as those first discussed in \cite{BBS92}, 
we recapitulate them to make this paper self-contained.
The details of the analysis in the Yukawa models at 
finite $T$ are found, for example, in \cite{BBS92} 
and \cite{KKN06,Kitazawa:2007ep}.

We start from the Lagrangian of a Yukawa model,
\begin{align}
{\cal L} 
= \bar\psi ( i\partial \hspace{-0.5em} / - m ) \psi
+ \frac12 \partial_\mu \phi \partial^\mu \phi 
- g_{\rm y} \phi \bar\psi\psi,
\label{eq:L_Yukawa}
\end{align}
where $\psi$ and $\phi$ denote the fermion and 
boson operators, $m$ is the fermion mass, and $g_{\rm y}$ 
represents the Yukawa coupling.
We neglect the mass term of the scalar boson,
since the purpose of this analysis is a study of 
quarks coupled to massless gauge bosons.
It is argued in \cite{BBS92} that the qualitative result 
about the fermion spectral function 
hardly changes even if we promote the scalar field in
Eq.~(\ref{eq:L_Yukawa}) to the $U(1)$ gauge field.
In the following, we call the fermion field $\psi$ 
the quark.

The quark self-energy in the imaginary-time formalism 
at one-loop order is given by,
\begin{align}
&\tilde\Sigma( i\omega_m,\bm{p} )
\nonumber \\
&= -g_{\rm y}^2 T \sum_n \int \frac{ d^3k }{ (2\pi)^3 }
{\cal S}_0 ( i\omega_n,\bm{k} ) 
{\cal D}_0 ( i\omega_m-i\omega_n,\bm{p}-\bm{k} ),
\end{align}
where 
${\cal S}_0 ( i\omega_n,\bm{p} ) 
= [ i\omega_n\gamma^0 - \bm{p}\cdot\bm{\gamma} - m ]^{-1}$ and
${\cal D}_0 ( i\nu_n,\bm{p} ) = [ (i\nu_n)^2 - \bm{p}^2 ]^{-1}$ 
are the Matsubara propagators
for the free quark and the free scalar boson, respectively,
with $\omega_n=(2n+1)\pi T$ and $\nu_n=2n\pi T$.
After summation over $n$ and analytic continuation, 
one obtains the self-energy in the real time, 
$\Sigma(\omega,\bm{p}) =\tilde\Sigma(i\omega_n,\bm{p})
|_{i\omega_n\to\omega}$.

The self-energy $\Sigma(\omega,\bm{p})$ has an ultraviolet
divergence, which can be removed with a standard 
renormalization \cite{BBS92,KKN06,Kitazawa:2007ep}.
Here we simply neglect the $T$-independent part, 
$\Sigma_{T=0}(\omega,\bm{p}) \equiv 
\lim_{T\to0} \Sigma(\omega,\bm{p})$, which includes 
the divergence.
This approximation is justified if the temperature is high
enough, since the $T$-dependent part, $\Sigma_{T\ne0}(\omega,\bm{p}) 
\equiv \Sigma(\omega,\bm{p}) - \Sigma_{T=0}(\omega,\bm{p})$,
grows rapidly and dominates over 
$\Sigma_{T=0}(\omega,\bm{p})$ as $T$ is raised.
The $T$-dependent part $\Sigma_{T\ne0}(\omega,\bm{p})$
does not suffer from any divergences and 
can be calculated without renormalization.
The spectral function at one-loop order is then given by,
\begin{align}
\rho( \omega,\bm{p} )
= -\frac1\pi {\rm Im} \frac1{ (\omega+i\eta)\gamma^0 
- \bm{p}\cdot\bm{\gamma} - m - \Sigma( \omega,\bm{p} ) }.
\label{eq:rho_Yukawa}
\end{align}

In our formalism, $m$ and $T$ are the only dimensionfull 
parameters and thus they control all properties of 
the system with a fixed Yukawa coupling $g_{\rm y}$.
In particular, the dimensionless spectral function,
$\tilde\rho = T\rho( \omega/T,\bm{p}/T )$, 
is determined uniquely for a given ratio $T/m$.
The limit $T/m\to0$ corresponds to low temperature, 
where $\rho(\omega,\bm{p})$ approaches the spectral function 
of free quarks, Eq.~(\ref{eq:rho_free}).
The opposite limit, $T/m\to\infty$, represents
the high temperature limit.
If we take $g_{\rm y}\to0$ in this limit, 
$\rho(\omega,\bm{p})$ becomes that calculated 
in the HTL approximation Eq.~(\ref{eq:rho^P_HTL}) 
with $m_T= g_{\rm y}T/4$.
With a fixed nonzero $g_{\rm y}$, the $\delta$-functions 
in $\rho( \omega,\bm{p} )$ become peaks having 
a non-zero width of order $g_{\rm y}^2T$.
One can, however, check numerically that the 
qualitative structure of $\rho( \omega,\bm{p} )$ hardly 
changes with a small Yukawa coupling 
$g_{\rm y}\lesssim1$.

\begin{figure}[tbp]
\begin{center}
\includegraphics[width=.49\textwidth]{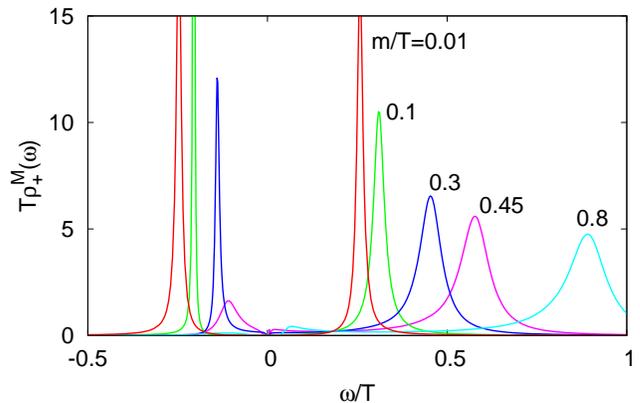}
\caption{
Spectral function $\rho^{\rm M}_+(\omega)$ 
in the Yukawa model Eq.~(\ref{eq:L_Yukawa}) for 
several values of $m/T$ with a Yukawa coupling 
$g_{\rm y}=1$.
}
\label{fig:Yukawa}
\end{center}
\end{figure}

To compare the spectral function in the Yukawa model 
with the results in Sec.~\ref{sec:T>Tc}, we limit 
our attention to zero momentum.
In this case, 
$\rho(\omega,\bm{0})$ is decomposed
as in Eq.~(\ref{eq:rho^M}) with projection operators $L_\pm$.
In the following we also regard $m/T$ as the external 
parameter, instead of $T/m$, since this is much more
convenient for the comparison with the lattice result.
From the above argument, one expects that the spectral 
function $\rho^{\rm M}_+(\omega)$ in the limit $m/T\to\infty$ 
approaches the free quark form, Eq.~(\ref{eq:rho^M_free}),
\begin{align}
\lim_{m/T\to\infty} 
\rho^{\rm M}_+(\omega) \simeq \delta( \omega - m ),
\label{eq:rho_Y0}
\end{align}
while in the opposite limit $m/T\to0$, $\rho^{\rm M}_+(\omega)$ 
should approach Eq.~(\ref{eq:rho^M_HTL}),
\begin{align}
\lim_{m/T\to0} 
\rho^{\rm M}_+(\omega) \simeq \frac12\left( 
\delta( \omega - m_T ) + \delta( \omega + m_T ) \right).
\label{eq:rho_Yinf}
\end{align}

We show the numerical result for $\rho^{\rm M}_+(\omega)$ 
for several values of $m/T$ in Fig.~\ref{fig:Yukawa}.
A fixed Yukawa coupling $g_{\rm y}=1$ is employed 
in this calculation:
We have checked that the qualitative feature of the 
numerical result does not change with a variation 
of $g_{\rm y}$ over a rather wide range.
The figure shows that $\rho^{\rm M}_+(\omega)$ for $m/T=0.01$ 
qualitatively reproduces Eq.~(\ref{eq:rho_Yinf}), 
having two peaks around $\omega=\pm gT/4$.
As $m/T$ increases, the peak at negative energy 
corresponding to the plasmino ceases to exist,
and $\rho^{\rm M}_+(\omega)$ is eventually dominated by 
a single peak at positive energy $\omega\simeq m$.
Although in the figure the width of the peak at positive energy,
$\Gamma$, grows as $m/T$ increases, one can check numerically 
and analytically that $\Gamma/m$ vanishes 
in the limit $m/T\to\infty$.
The width of the peak therefore is negligible in this limit, 
and $\rho^{\rm M}_+(\omega)$ reproduces Eq.~(\ref{eq:rho_Y0}).

The numerical result presented in Fig.~\ref{fig:Yukawa} 
shows that the two limits given by Eqs.~(\ref{eq:rho_Y0}) 
and (\ref{eq:rho_Yinf}), respectively, are connected 
continuously at the one-loop order.
It is also seen that the absolute value of the position of 
the peak at positive (negative) energy is a monotonically 
increasing (decreasing) function of $m/T$.
As discussed in the text, this feature is qualitatively 
different from that observed on the lattice near but above $T_c$.

\section{Exceptional configurations}
\label{sec:exceptional}

As mentioned in Sec.~\ref{sec:setup},
we found that the quark correlation functions 
$S(\tau,\bm{p})$ on some gauge configurations 
for $T/T_c \le 1.5$ behave anomalously near the 
chiral limit and at zero momentum.
In this appendix, we summarize the behavior of 
$S(\tau,\bm{p})$ on such exceptional configurations (EC).
A criterion to determine the EC used in the present 
analysis is also described.

\begin{figure}[tbp]
\begin{center}
\includegraphics[width=.49\textwidth]{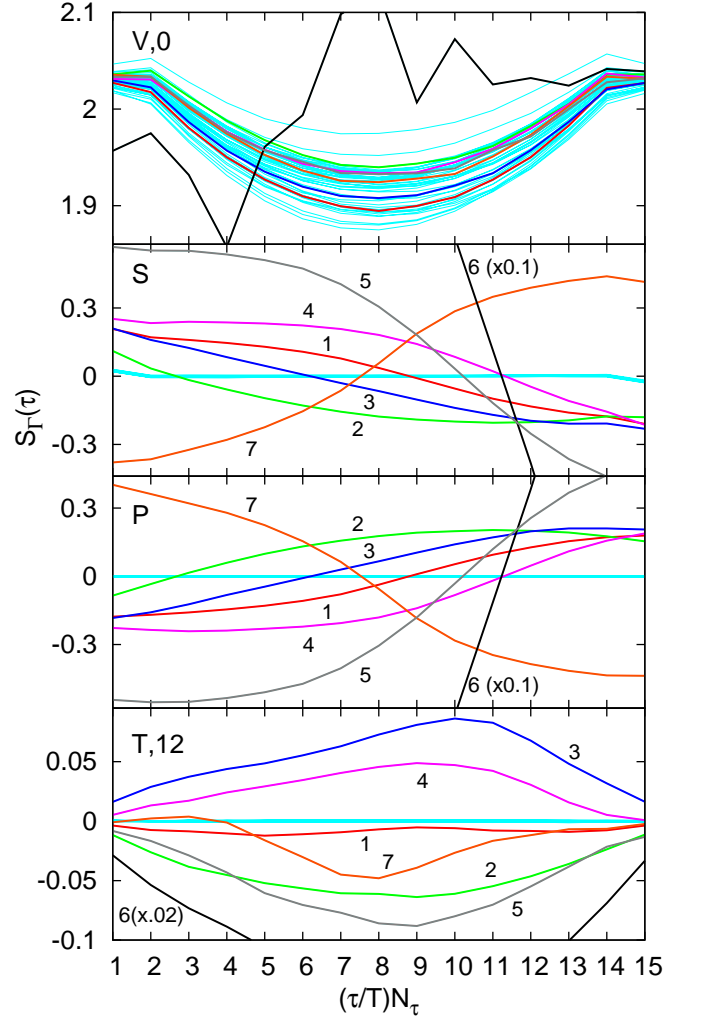}
\caption{
Quark correlation functions in the vector, scalar, 
pseudo-scalar and tensor channels, $S_{\rm V,0}(\tau)$, 
$S_{\rm S}(\tau)$, $S_{\rm P}(\tau)$ and $S_{\rm T,12}(\tau)$ 
(from top to bottom) on all configuration for a lattice
of size $64^3\times16$ for $T/T_c=1.5$.
The exceptional configurations are shown by the bold lines.
}
\label{fig:excp}
\end{center}
\end{figure}

For the moment, we consider the correlation function at zero 
momentum in the chiral limit $\kappa=\kappa_c$ on the lattice 
of size $64^3\times16$ for $T/T_c=1.5$ ($\beta=6.872$).
The Dirac structure of $S(\tau,\bm{0})$ is decomposed as
\begin{align}
S_{\rm S}( \tau ) 
&= {\rm Tr}_{\rm D} \left[ S(\tau,\bm{0}) \right]/4, 
\label{eq:S_s}\\
S_{\rm V,\mu}( \tau ) 
&= {\rm Tr}_{\rm D} \left[ \gamma_\mu S(\tau,\bm{0}) \right]/4, \\
S_{\rm T,\mu\nu}( \tau ) 
&= {\rm Tr}_{\rm D} \left[ \sigma_{\mu\nu} S(\tau,\bm{0}) \right]/4, \\
S_{\rm A,\mu}( \tau ) 
&= {\rm Tr}_{\rm D} \left[ \gamma_\mu \gamma_5 S(\tau,\bm{0}) \right]/4, \\
S_{\rm P}( \tau ) 
&= {\rm Tr}_{\rm D} \left[ \gamma_5 S(\tau,\bm{0}) \right]/4.
\label{eq:S_p}
\end{align}
In Fig.~\ref{fig:excp}, we show 
$S_{\rm V,0}(\tau)$, $S_{\rm S}(\tau)$, 
$S_{\rm P}(\tau)$, and $S_{\rm T,12}(\tau)$
on all $51$ configurations. 
Among them, seven configurations are specified 
as the EC which are depicted by the bold lines. 
The numbers $1$-$7$ are also assigned to these 
lines for better identification of each configuration.
The correlation functions obtained on the other $44$ 
configurations are shown by thin light-blue lines, which 
are, however, almost degenerated in the lower three panels.
From the figure, one clearly sees that the behavior of 
$S_{\rm S}(\tau)$, $S_{\rm P}(\tau)$, and 
$S_{\rm T,12}(\tau)$ on the EC is qualitatively 
different from other normal configurations which are 
approximately zero in these channels.
As discussed in Sec.~\ref{sec:propagator}, 
the chiral, parity, and rotational symmetries require 
that the correlation functions in these channels vanish.
The behavior of these functions on the EC therefore is 
obviously unphysical.
On the other hand, $S_{\rm V,0}(\tau)$ tends to behave 
moderately even on the EC.

\begin{figure}[tbp]
\begin{center}
\includegraphics[width=.49\textwidth]{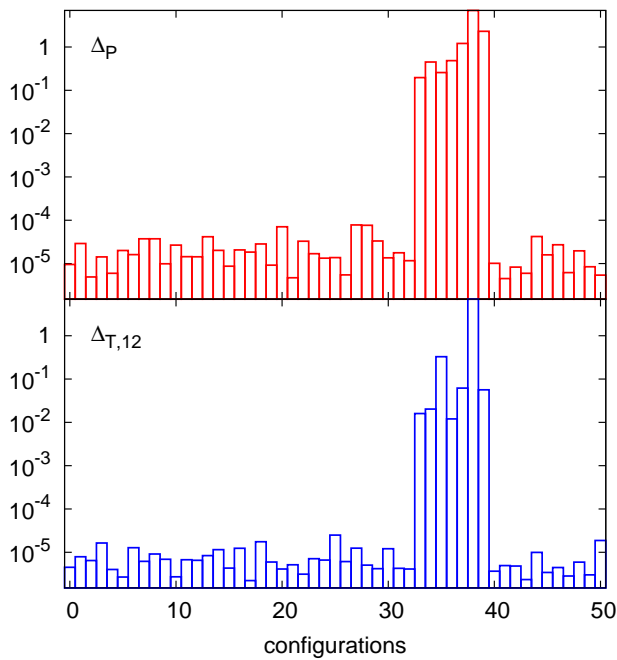}
\caption{
Values of $\Delta_{\rm P}$ and $\Delta_{\rm T,12}$
on all configurations for a lattice of size
$64^3\times16$ and $T/T_c=1.5$.
}
\label{fig:Delta}
\end{center}
\end{figure}

Since the anomalous behavior of $S_{\rm S}(\tau)$, 
$S_{\rm P}(\tau)$, and $S_{\rm T,12}(\tau)$ on 
the EC is quite evident,
it is easy to introduce a criterion to identify the EC.
Here, we introduce 
\begin{align}
\Delta_\Gamma = \sum_{\tau=\tau_{\rm min}}^{N_\tau-\tau_{\rm min}} 
|S_\Gamma (\tau)|^2,
\end{align}
for each configuration where $\Gamma$ defines different 
channels Eqs.~(\ref{eq:S_s}) - (\ref{eq:S_p}),
and regard the configurations satisfying
\begin{align}
\Delta_\Gamma > D,
\label{eq:Delta}
\end{align}
as the exceptional ones with $D$ being the threshold 
to be determined empirically.
We show $\Delta_{\rm P}$ and $\Delta_{\rm T,12}$
with $\tau_{\rm min}=3$ on all configurations 
in Fig.~\ref{fig:Delta}.
The horizontal axis represents the different gauge 
configurations which are ordered 
according to Monte Carlo time.
One sees that $\Delta_{\rm P}$ and $\Delta_{\rm T,12}$
on the EC take values more than two orders of magnitude 
larger than the typical ones on the normal configurations.
This result means that there is a wide range 
for the choice of $D$ in Eq.~(\ref{eq:Delta}),
and hence this criterion works well in practice.
Our numerical result shows that the criterion 
Eq.~(\ref{eq:Delta}) is most successfully applied
to the pseudoscalar channel, $\Delta_{\rm P}$,
and successively $\Delta_{{\rm T},ij}$ with $1\le i,j\le3$.
Here, we notice that our formula for $S(\tau,\bm{p})$
with the wall source Eq.~(\ref{eq:wall}), instead of 
Eq.~(\ref{eq:point}), plays a crucial role for this 
clear separation between the normal and exceptional 
configurations.
In fact, with the correlation function calculated 
with Eq.~(\ref{eq:point}), the quark correlation 
functions have large fluctuations and 
the range of $D$ which separates the EC with
Eq.~(\ref{eq:Delta}) becomes narrower.

\begin{figure}[tbp]
\begin{center}
\includegraphics[width=9cm]{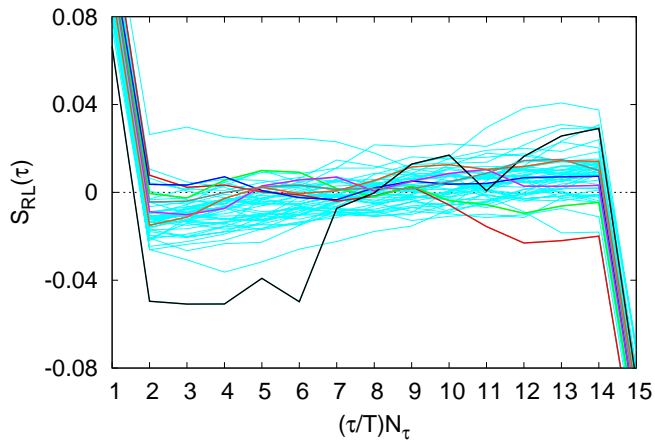}
\caption{
Behavior of the correlation functions 
$S_{\rm RL}(\tau)$ on all configurations for a 
lattice of size $64^3\times16$ and $T/T_c=1.5$.
The exceptional configurations are shown by the bold lines.
}
\label{fig:excp_lr}
\end{center}
\end{figure}

Figure~\ref{fig:Delta} also shows that the EC on this set of 
gauge configurations are strongly correlated. 
They correspond to subsequent configurations in Monte Carlo 
time, although the separation between these gauge 
configurations is a few times larger than the autocorrelation
length measured in terms of the plaquette and Polyakov loop 
correlation functions \cite{Bielefeld}.
This shows that the correlation among EC is significantly 
stronger and leads to a much larger autocorrelation length.
A similar result is obtained for $T/T_c=1.25$, although in 
this case we observed several groups of such successive EC.

Next, in Fig.~\ref{fig:excp_lr} we show the correlation function
representing the propagation between right and left handed
quarks, $\psi_{\rm R,L} = (1/2)(1\pm\gamma_5)\psi$,
\begin{align}
S_{\rm RL}(\tau) 
&= \langle {\rm T}_\tau \psi_{\rm R}(\tau) 
\bar\psi_{\rm L}(0) \rangle
= {\rm Tr}_{\rm D} [ \frac{1+\gamma_5}2 S(\tau,\bm{0}) ] 
\nonumber \\
&= \frac12 ( S_{\rm S}(\tau)+S_{\rm P}(\tau) ).
\end{align}
The figure shows that $S_{\rm RL}(\tau)$ is close to zero as
it should be even on the EC.
On the other hand, the opposite channel 
$S_{\rm LR}(\tau) = {\rm Tr}_{\rm D}[S(\tau,\bm{0}) (1-\gamma_5)/2]
=(S_{\rm S}(\tau)-S_{\rm P}(\tau))/2$
behaves anomalously on the EC, as expected from the behavior
of $S_{\rm S,P}(\tau)$ in Fig.~\ref{fig:excp}.
Although on the gauge configurations for $T/T_c=1.5$ 
we observed the anomalous behavior only on $S_{\rm LR}(\tau)$
for all configurations, we checked that for $T/T_c=1.25$ 
there appear anomalous behaviors in both 
$S_{\rm LR}(\tau)$ and $S_{\rm RL}(\tau)$.
For $T/T_c=1.25$, however, only one of them tends to 
behave anomalously on each EC with a few exceptions.
It is also found that the channel having the exceptional 
behavior tends to be common in a group of configurations
appearing successively.
The results presented above indicate that there exist 
topological objects on the EC that cause the anomalous 
behavior of the quark correlation functions.
To check this speculation, it would be interesting to directly 
measure the topological charge on each configuration. 

Finally, we remark on a relation between 
EC observed in the quark correlation function
and those in the hadronic channels.
By measuring the pion correlation function
constructed from the quark correlation function
with the point source, Eq.~(\ref{eq:point}), 
we confirmed that the appearance of anomalous behavior
in the pion correlation function is limited on the EC 
determined with the criterion Eq.~(\ref{eq:Delta}).
We, however, also found that the behavior of pion 
propagator {\it seems} moderate on some configurations 
satisfying Eq.~(\ref{eq:Delta}).
The latter behavior may be attributed to the 
form of the lattice correlation function:
As mentioned above, the wall source, Eq.~(\ref{eq:wall}), 
for the quark propagator drastically reduces the statistical
fluctuations compared to the point source, and hence 
allows the clear separation of EC with
a criterion like  Eq.~(\ref{eq:Delta}).
For the same reason, the fluctuations in the pion channel
can be large when calculated with a point source and such 
fluctuations make the identification of the EC difficult.
We thus expect that if we would calculate the pion correlation 
function with a wall source operator similar to that used 
in Eq.~(\ref{eq:wall}) for the quarks, there would be 
a perfect correspondence in the appearance of the 
exceptional behavior in both correlation functions.

\end{document}